\documentclass{article}
\usepackage{PRIMEarxiv}
\usepackage[utf8]{inputenc} 
\usepackage[T1]{fontenc}    
\usepackage{hyperref}       
\usepackage{url}            
\usepackage{booktabs}       
\usepackage{amsfonts}       
\usepackage{nicefrac}       
\usepackage{microtype}      
\usepackage{lipsum}
\usepackage{fancyhdr}       
\usepackage{graphicx}       
\usepackage{amsmath,amsfonts}
\usepackage{booktabs}
\usepackage{enumerate}
\graphicspath{{media/}}     

\title{Unified Error Correction Code Transformer with Low Complexity}

\author{
  Yongli Yan \\
  Department of Electronic Engineering \\
  Tsinghua University \\
  \texttt{yanyongli@tsinghua.edu.cn} \\
  \And
  Jieao Zhu, Tianyue Zheng \\
  Department of Electronic Engineering \\
  Tsinghua University \\
  \texttt{{zja21, zhengty22}@mails.tsinghua.edu.cn} \\
  \And
  Zhuo Xu, Chao Jiang \\
  Department of Electronic Engineering \\
  Tsinghua University \\
  \texttt{{xz23, jiangc24}@mails.tsinghua.edu.cn} \\
  \And
  Linglong Dai \\
  Department of Electronic Engineering \\
  Tsinghua University \\
  \texttt{daill@tsinghua.edu.cn} \\
}

\newcounter{magicrownumbers}
\newcommand\rownumber{\stepcounter{magicrownumbers}\arabic{magicrownumbers}}

\begin{document}
\maketitle

\begin{abstract}
Channel coding is vital for reliable sixth-generation (6G) data transmission, employing diverse error correction codes for various application scenarios. Traditional decoders require dedicated hardware for each code, leading to high hardware costs. Recently, artificial intelligence (AI)-driven approaches, such as the error correction code Transformer (ECCT) and its enhanced version, the foundation error correction code Transformer (FECCT), have been proposed to reduce the hardware cost by leveraging the Transformer to decode multiple codes. However, their excessively high computational complexity of $\mathcal{O}(N^2)$ due to the self-attention mechanism in the Transformer limits scalability, where $N$ represents the sequence length. To reduce computational complexity, we propose a unified Transformer-based decoder that handles multiple linear block codes within a single framework. Specifically, a standardized unit is employed to align code length and code rate across different code types, while a redesigned low-rank unified attention module, with computational complexity of $\mathcal{O}(N)$, is shared across various heads in the Transformer. Additionally, a sparse mask, derived from the parity-check matrix's sparsity, is introduced to enhance the decoder's ability to capture inherent constraints between information and parity-check bits, improving decoding accuracy and further reducing computational complexity by $86\%$. Extensive experimental results demonstrate that the proposed unified Transformer-based decoder outperforms existing methods and provides a high-performance, low-complexity solution for next-generation wireless communication systems.
\end{abstract}

\section{Introduction}
With the continuous evolution of wireless communication technologies, the imminent arrival of sixth-generation (6G) wireless communication will mark a paradigm shift toward ultra-reliable communication~\cite{massive_mimo_survey, toward_6g, road_to_6g, channel_coding_6g}. To meet the stringent performance demands of 6G, channel codes with strong error correction capabilities are expected to play a pivotal role. Specifically, low-density parity-check (LDPC) codes~\cite{gallager_ldpc} and Polar codes~\cite{arikan_polar_code} have proven effective in fifth-generation (5G) wireless communications. Due to their excellent performance, 3GPP has just made the decision to continue to use LDPC and Polar codes for future 6G in October 2025~\cite{ran1_coding_std}. Furthermore, algebraic geometry codes, such as Bose-Chaudhuri-Hocquenghem (BCH) codes~\cite{bch_code}, are commonly employed as outer codes in concatenated schemes with probabilistic codes to enhance decoding performance.

\subsection{Prior Works}
In wireless communication systems, multiple coding schemes are supported, such as LDPC and Polar codes in 5G~\cite{3gpp_38212_std}, each requiring customized hardware for their decoding algorithms. For instance, LDPC codes employ iterative message passing for \textbf{\textit{parallel}} data processing~\cite{hardware_ldpc}, while Polar codes rely on the successive cancellation (SC) decoding algorithm~\cite{arikan_polar_code}, which follows a \textbf{\textit{sequential}} approach by traversing a decision tree, often simplified through binary tree pruning~\cite{hardware_sc, hardware_bp, hardware_reduced, hardware_scl}. Both LDPC and Polar codes utilize \textbf{\textit{soft}} decoding methods, leveraging probabilistic information to achieve robust error correction. In contrast, BCH codes, another key coding scheme, employ algebraic decoding techniques classified as \textbf{\textit{hard}} decoding~\cite{hardware_bch}, which rely on mathematical structures to correct errors.The gap between the hard decoding of BCH codes and the soft decoding of LDPC and Polar codes, combined with the distinct parallel and sequential architectures of LDPC and Polar codes, necessitates customized hardware circuits for each code type, resulting in significant hardware resource cost.

To reduce hardware resource cost, three approaches have been investigated for decoding multiple codecs with a single hardware framework: resource sharing, ordered statistical decoding, and artificial intelligence (AI)-driven methods. First, resource-sharing methods, such as those in~\cite{pipelined_arch, unified_fec}, integrate LDPC and Polar code decoders into a single architecture using traditional decoding algorithms. These methods conserve hardware resources by multiplexing shared computational and storage units through reconfiguration. However, due to the distinct decoding algorithms for each code type, only partial resource sharing is achieved, limiting hardware resource reduction.

Second, ordered-statistics decoding methods, as explored in~\cite{osd_decoder0, osd_decoder1, osd_decoder2}, provide near-maximum likelihood (ML) decoding for linear block codes by ordering received symbols by reliability, enabling hardware reuse and reducing resource costs. Ordered-statistics decoding sorts signal positions by reliability, permutes the generator matrix of the code in a consistent manner across different linear block codes, and evaluates the most likely codewords through a limited search. This uniform approach to generator matrix processing enables hardware reuse across various linear block codes, reducing resource cost compared to code-specific decoding architectures. However, the computational complexity of ordered-statistics decoding increases exponentially with code length and search order (e.g., order-$l$). For example, an order-4 ordered-statistics decoding may require evaluating thousands of test codewords, increasing decoding latency.

Third, in recent years, AI-driven decoders have gained attention as an alternative to reduce hardware resource cost. For example, inspired by the success of Transformers in natural language processing~\cite{attn_2017}, the error correction code Transformer (ECCT) was introduced to reduce hardware resource and enhance decoding performance for short codes~\cite{ecct_2022}. The inherent connection between Transformers and decoders lies in their shared capability for sequence-to-sequence transformations, mapping input sequences—e.g., from one language to another in natural language processing or from log-likelihood ratios (LLRs) to information bits in channel decoding—into output sequences. ECCT leverages the Transformer’s self-attention mechanism, where multiple attention heads operate in parallel to process sequential data and capture long-range dependencies, effectively modeling the constrained relationships among the LLRs of decoder inputs. Building on ECCT, the foundation error correction code Transformer (FECCT) enhances generalization to unseen codewords by integrating the distance matrix, derived from the parity-check matrix $\mathbf{H}$, into the self-attention mechanism~\cite{fecct_2024}. However, ECCT and FECCT rely on the Transformer’s powerful self-attention mechanism, which has a computational complexity of $\mathcal{O}(N^2)$, where $N$ represents the sequence length. Their high computational complexity makes them difficult to scale to long codes.

\subsection{Contributions}

To reduce computational complexity, we propose a unified Transformer-based decoder capable of handling multiple linear block codes, including LDPC, Polar, and BCH, within a single framework. To achieve this, we design a standardized unit and a low-rank unified attention module that shares attention scores across various heads in the Transformer. Additionally, we develop a sparse masking mechanism to enhance decoding performance and further reduce computational complexity\footnote{Simulation codes will be provided to reproduce the results in this paper: \url{http://oa.ee.tsinghua.edu.cn/dailinglong/publications/publications.html.}}. Specifically, the main contributions of this paper are summarized as follows.

\begin{enumerate}
\item We introduce a unified Transformer-based decoder capable of seamlessly integrating various linear block codes. This architecture incorporates a standardized unit to align code lengths, a unified attention module that compresses the structural information of all codewords, and a sparse mask leveraging the parity-check matrix's sparsity to enhance decoding accuracy. These components achieve a code-agnostic decoder, eliminate the need for distinct hardware circuits for different decoding algorithms, and reduce computational complexity.

\item A standardized unit aligns parameters, such as code length and code rate, across different code types. Specifically, it pads codewords and their corresponding syndromes with zeros to a maximum length after receiving the channel output. This enables the Transformer-based decoder to handle any codeword within the maximum length, facilitating unified joint training.

\item We redesigned the self-attention module to create a unified attention mechanism that allows different attention heads to share attention scores. This approach contrasts with traditional self-attention mechanisms, where each head independently computes its scores using separate learned weights. This unified attention compresses structural information of different codewords through learning and shares parameters across various heads in the Transformer. It eliminates the need to project inputs into higher dimensions, reducing computational complexity from $\mathcal{O}(N^2)$ to $\mathcal{O}(N)$, where $N$ represents the sequence length. This results in a more efficient and unified decoding architecture.

\item By introducing a sparse mask, which leverages the sparsity of the parity-check matrix, we have significantly enhanced the model's ability to capture inherent constraints between information and parity-check bits. This has led to a substantial improvement in decoding performance and the achievement of a state-of-the-art design. Additionally, this sparse mask further reduces computational complexity by $86\%$, enabling efficient training for long codes.
\end{enumerate}

The rest of this paper is organized as follows. Section II provides background knowledge, including the vanilla Transformers, channel encoders, as well as pre- and post-processing involved to address the issue of overfitting. Section III details our proposed unified Transformer-based decoder. Section IV presents the results of training and inference. Section V analyzes the impact of the proposed optimization algorithms. Finally, Section VI concludes.

\textit{Notations}: Bold lowercase and uppercase letters denote vectors (e.g., $\mathbf{v}$) and matrices (e.g., $\mathbf{M}$), respectively, while scalars use regular letters (e.g., $x$). $\mathbb{F}_2$ represents the Galois field of order 2, and $\mathbb{R}$, $\mathbb{N}$ denote the real and natural numbers. The transpose is indicated by $[\cdot]^T$, and $\mathcal{N}(\mu, \sigma^2)$ denotes a Gaussian distribution with mean $\mu$ and variance $\sigma^2$. The function $\text{bipolar}(x) = 1 - 2x$ maps $x \in {0, 1}$ to bipolar values ${1, -1}$, with its inverse $\text{bin}(x) = \frac{1 - x}{2}$ mapping ${1, -1}$ back to ${0, 1}$. The operator $\text{size}(\cdot)$ denotes the number of elements in a vector or matrix, and $\text{sign}(\cdot)$ represents the sign function, defined as $\text{sign}(x) = 1$ if $x > 0$, $0$ if $x = 0$, and $-1$ if $x < 0$. For a linear block code with code length $n$ and information bit length $k$, we denote it as $(n, k)$.

\section{Background}
For a better understanding of the proposed unified error correction code Transformer, this section will provide background knowledge on vanilla Transformers and forward error correction (FEC) codes.

\subsection{Vanilla Transformers}
\begin{figure}[htb]
\centering
\includegraphics[width=0.3\columnwidth]{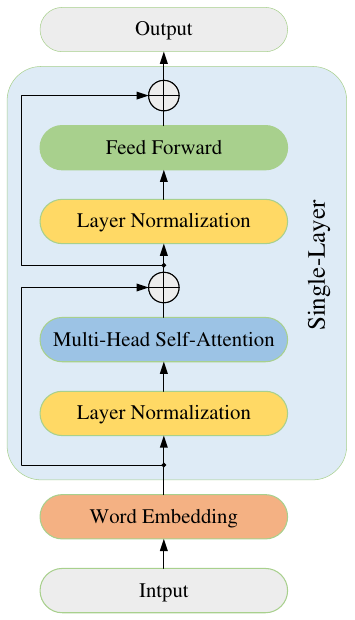}
\caption{Single-layer vanilla Transformer encoder architecture.}
\label{fig_arch_transformer}
\end{figure}

The vanilla Transformer, a sequence-to-sequence model, comprises an encoder and decoder, each with $L$ identical blocks. Each block integrates multi-head self-attention (MHA), a position-wise feed-forward network (FFN), residual connections to mitigate gradient issues~\cite{cvpr_2016}, and layer normalization  for stability and faster convergence~\cite{layer_norm_2016}, as shown in Figure \ref{fig_arch_transformer}.

The process begins with embedding a one-dimensional input signal $\textbf{\textit{x}} \in \mathbb{R}^{N}$ into a higher-dimensional space:
\begin{equation}
\label{eq_emb}
\begin{aligned}
\textbf{\textit{X}}_{e} = \text{Embedding}(\textbf{\textit{x}}) = \text{diag}(\textbf{\textit{x}}) \cdot \textbf{\textit{W}},
\end{aligned}
\end{equation}
where $\text{diag}(\textbf{\textit{x}}) \in \mathbb{R}^{N \times N}$ is a diagonal matrix with $\textbf{\textit{x}}$ on the diagonal, and $\textbf{\textit{W}} \in \mathbb{R}^{N \times d_k}$ is a trainable weight matrix, producing $\textbf{\textit{X}}_e \in \mathbb{R}^{N \times d_k}$ with $d_k$-dimensional embeddings for each position.

\begin{figure}[htb]
\centering
\includegraphics[width=0.7\columnwidth]{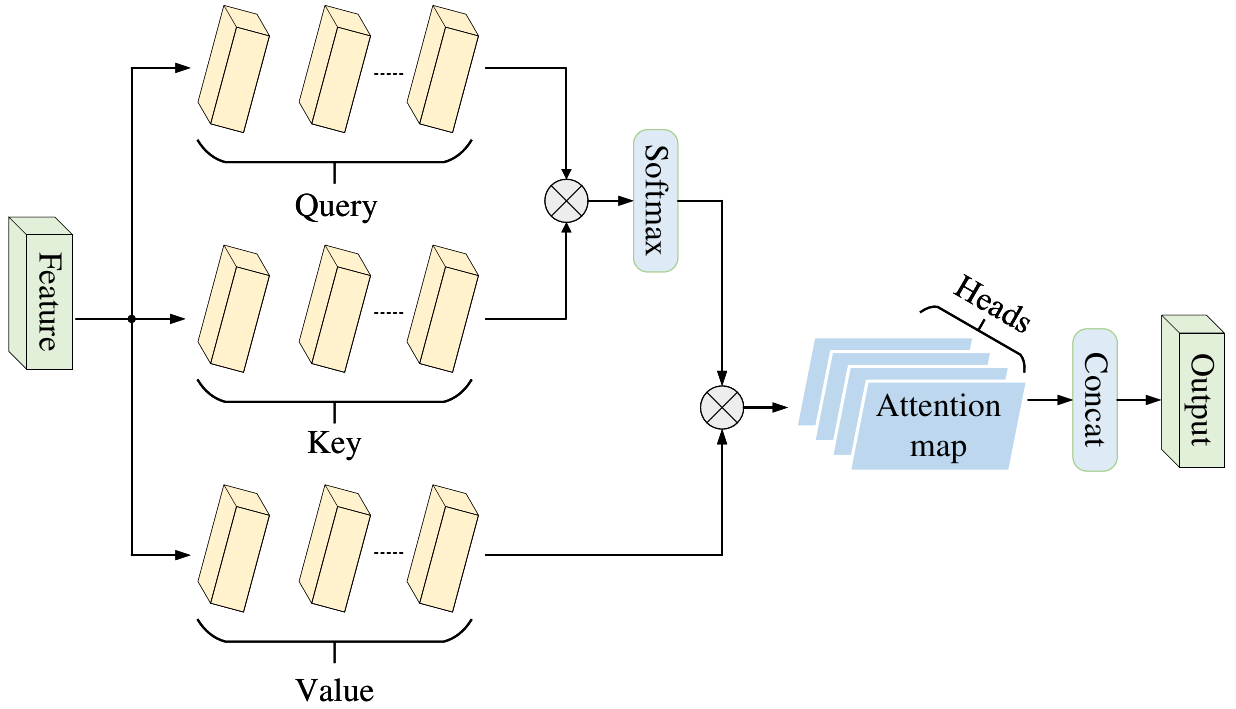}
\caption{Architecture of multi-head self-attention.}
\label{fig_arch_mhsa}
\end{figure}

This embedding feeds into the MHA mechanism, which correlates sequence elements through:
\begin{equation}
\label{eq_attn}
\begin{aligned}
\text{Attn}\left(\textbf{\textit{Q}},\textbf{\textit{K}},\textbf{\textit{V}}\right)=\text{Softmax}\left(\frac{\textbf{\textit{Q}} \cdot \textbf{\textit{K}}^T}{\sqrt{d_k}}\right) \cdot \textbf{\textit{V}},
\end{aligned}
\end{equation}
extended to multiple heads as:
\begin{equation}
\label{eq_mha}
\begin{aligned}
\text{MHA}\left(\textbf{\textit{Q}},\textbf{\textit{K}},\textbf{\textit{V}}\right)=
\text{Concat}\left(\text{head}_1,\cdots,\text{head}_H\right) \cdot \textbf{\textit{W}}^\textit{O},
\end{aligned}
\end{equation}
where $\text{head}_i = \text{Attn}(\textbf{\textit{X}}_Q \textbf{\textit{W}}_i^Q, \textbf{\textit{X}}_K \textbf{\textit{W}}_i^K, \textbf{\textit{X}}_V \textbf{\textit{W}}_i^V)$, with $\textbf{\textit{X}}_Q, \textbf{\textit{X}}_K, \\ \textbf{\textit{X}}_V \in \mathbb{R}^{N \times d_k}$ set to $\textbf{\textit{X}}_e$ for self-attention, using trainable weights $\textbf{\textit{W}}_i^Q, \textbf{\textit{W}}_i^K, \textbf{\textit{W}}_i^V \in \mathbb{R}^{d_k \times d_k}$ and $\textbf{\textit{W}}^O \in \mathbb{R}^{H d_k \times H d_k}$, as depicted in Figure \ref{fig_arch_mhsa}.

Next, the MHA output is processed by the FFN, applied uniformly across positions:
\begin{equation}
\label{eq_ffn}
\begin{aligned}
\text{FFN}\left( \textbf{\textit{X}} \right)=\textbf{\textit{W}}_2 \cdot \text{ReLU}\left(\textbf{\textit{W}}_1 \cdot \textbf{\textit{X}} + \textbf{\textit{b}}_1\right) + \textbf{\textit{b}}_2,
\end{aligned}
\end{equation}
where $\textbf{\textit{W}}_1 \in \mathbb{R}^{d_f \times d_h}$, $\textbf{\textit{W}}_2 \in \mathbb{R}^{d_h \times d_f}$, $\textbf{\textit{b}}_1$, $\textbf{\textit{b}}_2$ are trainable parameters, $d_h=H \cdot d_k$ and $\text{ReLU}(x) = \max(0, x)$ enables complex feature learning, with $d_f > d_h$ for enhanced performance.

Finally, residual connections and layer normalization link the MHA and FFN layers to produce the output $\textbf{\textit{X}}_2$ of a single-layer Transformer encoder:
\begin{equation}
\label{eq_transformer}
\begin{aligned}
\textbf{\textit{X}}_1&=\text{MHA}\left(\text{LayerNorm}\left(\textbf{\textit{X}}_e\right)\right) + \textbf{\textit{X}}_e \\
\textbf{\textit{X}}_2&=\text{FFN}\left(\text{LayerNorm}\left(\textbf{\textit{X}}_1\right)\right) + \textbf{\textit{X}}_1.
\end{aligned}
\end{equation}

\subsection{Forward Error Correction Codes}
In information theory, FEC is a technique designed to reduce the bit error rate (BER) when transmitting data over a noisy channel. We adopt a linear block code defined by a generator matrix $\textbf{G} \in \mathbb{R}^{k \times n}$ and a parity-check matrix $\textbf{H} \in \mathbb{R}^{(n-k) \times n}$, satisfying $\textbf{G} \cdot \textbf{H}^T = \textbf{0}$ over $\mathbb{F}_2$.

The encoder maps a message vector $\textbf{\textit{m}} \in \{0,1\}^k$ to a codeword $\textbf{\textit{x}} \in \{0,1\}^n$ via $\textbf{\textit{x}} = \textbf{\textit{m}} \cdot \textbf{G}$, where $\textbf{H} \cdot \textbf{\textit{x}} = \textbf{0}$. This codeword is modulated using binary phase shift keying (BPSK), defined as $\text{BPSK}(x) = 1 - 2x$, yielding $\textbf{\textit{x}}_s$, which is transmitted over binary-input discrete memoryless channels (B-DMCs, e.g., additive white Gaussian noise (AWGN) channels), producing $\textbf{\textit{y}} = \textbf{\textit{x}}_s + \textbf{\textit{n}}$, with $\textbf{\textit{n}} \sim \mathcal{N}(0, \sigma^2)$.

\begin{figure}[htb]
\centering
\includegraphics[width=0.8\columnwidth]{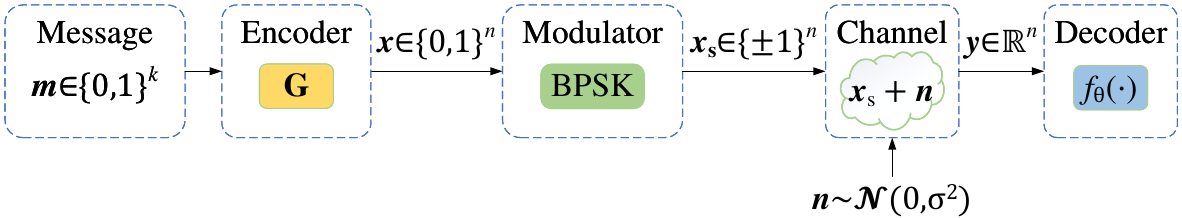}
\caption{Architecture of the communication system.}
\label{fig_arch_comm}
\end{figure}

\begin{figure}[htb]
\centering
\includegraphics[width=0.8\columnwidth]{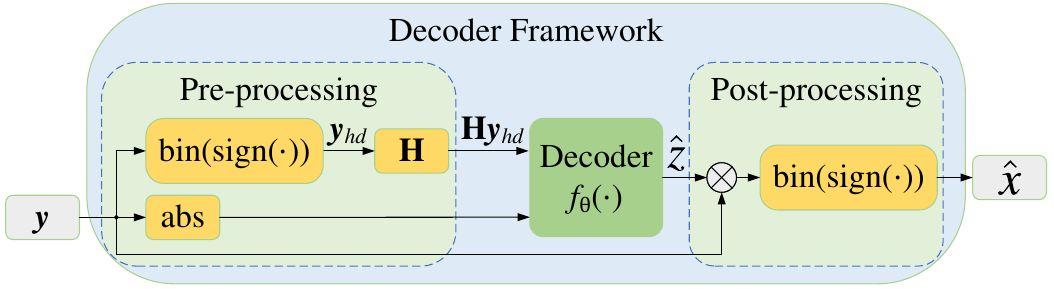}
\caption{Decoder framework with pre- and post-processing.}
\label{fig_arch_dec}
\end{figure}

The decoder aims to recover $\textbf{\textit{m}}$ from $\textbf{\textit{y}}$, as depicted in Figure \ref{fig_arch_comm}. Our work focuses on designing and training a parameterized decoding function $f_{\theta}$. To develop a data-driven Transformer-based decoder, we leverage pre- and post-processing techniques from~\cite{dl_decoding_syndrome}, preserving BER and mean squared error (MSE) performance (see Figure \ref{fig_arch_dec}). Training with an all-zero codeword mitigates overfitting due to exponential codeword space growth. And the B-DMCs is modeled by (\ref{eq_biso_mode}) which is an equivalent statistical mode that differs from the true physical one~\cite{capacity_ldpc_mp}.
\begin{equation}
\label{eq_biso_mode}
\begin{aligned}
\textbf{\textit{y}}=\textbf{\textit{x}}_s \cdot \textbf{\textit{z}},
\end{aligned}
\end{equation}
where ${\textbf{\textit{z}}}$ is a random multiplicative noise independent of ${\textbf{\textit{x}}_s}$.

In the pre-processing step, the absolute values of received signal and the syndromes are concatenated as the input to the neural network, which can be written as:
\begin{equation}
\label{eq_pre_process}
\begin{aligned}
\widetilde {\textbf{\textit{y}}}=\left[\lvert \textbf{\textit{y}} \rvert, \text{s}\left( \textbf{\textit{y}} \right) \right],
\end{aligned}
\end{equation}
where, $\left[ \cdot,\cdot \right]$ denotes the concatenation of vectors, $\lvert \textbf{\textit{y}} \rvert$ signifies the absolute value of $\textbf{\textit{y}}$ and $\text{s}(\textbf{\textit{y}}) = \textbf{H} \cdot \text{bin}(\text{sign}(\textbf{\textit{y}})) \in \{0,1\}^{n-k}$ is the syndrome from hard-decoded $\textbf{\textit{y}}$.

In the post-processing step, the predicted noise $\hat {\textbf{\textit{z}}}$ is multiplied by the received signal $\textbf{\textit{y}}$ to recover $\textbf{\textit{x}}$. That is, the predicted $\hat {\textbf{\textit{x}}}$ takes the form:
\begin{equation}
\label{eq_post_process}
\begin{aligned}
\hat {\textbf{\textit{x}}} = \text{bin} \left( \text{sign} \left( {\textbf{\textit{y}}} \cdot {\hat {\textbf{\textit{z}}}} \right) \right).
\end{aligned}
\end{equation}

\subsection{Mask Generation in ECCT and FECCT}
Masks in Transformer-based error correction enhance decoding by integrating code structure into attention mechanisms. In ECCT~\cite{ecct_2022}, the mask $g(\textbf{H}): \{0,1\}^{(n-k) \times n} \to \{-\infty, 0\}^{(2n-k) \times (2n-k)}$ embeds code structure using the parity-check matrix $\textbf{H}$. Initialized as an identity matrix, it unmasks positions of paired ones in $\textbf{H}$ rows, extending the Tanner graph to two-ring connectivity. It is applied additively in attention: $\text{Attn}(\textbf{\textit{Q}}, \textbf{\textit{K}}, \textbf{\textit{V}}) = \text{Softmax}\left(\frac{\textbf{\textit{Q}} \cdot \textbf{\textit{K}}^T + g(\textbf{H})}{\sqrt{d_k}}\right) \cdot \textbf{\textit{V}}$.

In FECCT~\cite{fecct_2024}, a learned $\psi: \mathbb{N} \to \mathbb{R}$ modulates attention via the distance matrix $\mathcal{G}(\textbf{H}) \in \mathbb{N}^{(2n-k) \times (2n-k)}$, applied with a Hadamard product: $\text{Attn}(\textbf{\textit{Q}}, \textbf{\textit{K}}, \textbf{\textit{V}}) = \left(\text{Softmax}\left(\frac{\textbf{\textit{Q}} \cdot \textbf{\textit{K}}^T}{\sqrt{d_k}}\right) \odot \psi(\mathcal{G}(\textbf{H}))\right) \cdot \textbf{\textit{V}}$.

\section{Proposed Unified Error Correction Code Transformer}
In this section, we present the detailed architecture and training process of the proposed \textbf{unified error correction code Transformer (UECCT)}.

\subsection{Shared Unified Attention}
The vanilla self-attention mechanism in Transformer architectures establishes dense correlations among all input features within a sequence, resulting in a computational complexity of $\mathcal{O}(N^2 \cdot d_k + N \cdot d_k^2)$, where $N$ is the sequence length and $d_k$ is the embedding dimension. This approach, while effective for capturing long-range dependencies, is computationally intensive and overlooks the intrinsic relationships across different samples, particularly in channel decoding tasks involving diverse linear block codes. Moreover, prior studies suggest that the input to the softmax function in self-attention is probabilistically sparse, with only a few weights dominating the attention distribution~\cite{transformer_long_seq}. This sparsity implies that a low-rank approximation could effectively capture the dominant correlations, reducing redundancy without compromising performance.

\begin{figure}[htb]
\centering
\includegraphics[width=0.6\columnwidth]{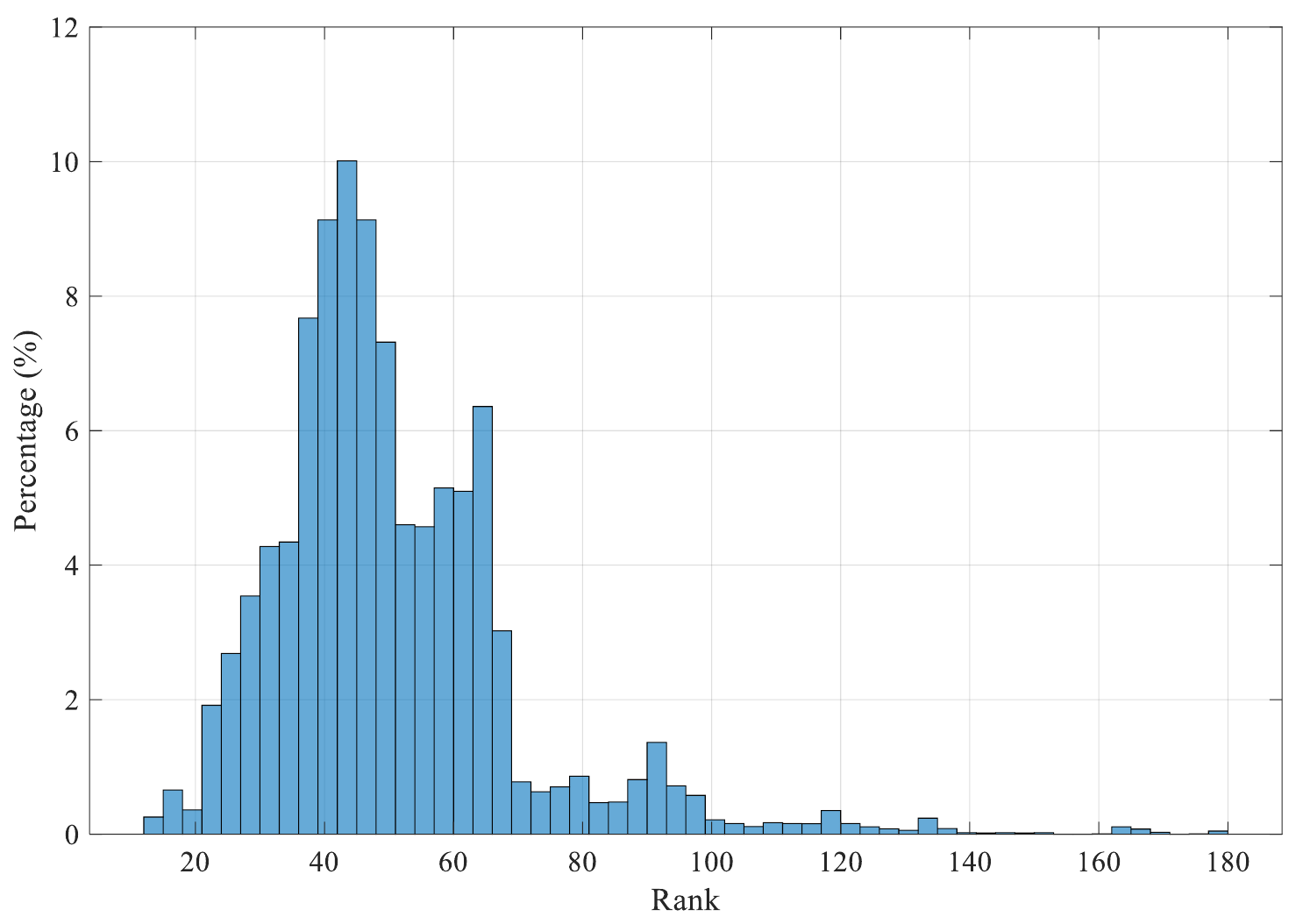}
\caption{Rank distribution of the $\textbf{\textit{Q}} \cdot \textbf{\textit{K}}^T$ matrix in the ECCT model across 6 encoder layers for various code types.}
\label{fig_ecct_rank}
\end{figure}

To investigate this hypothesis, we trained an ECCT model~\cite{ecct_2022} with parameters $L=6$ (number of Transformer encoder layers), $H=8$, $d_k=\text{max}(2n-k)=182$, and $d_f=2912$ on a diverse set of linear block codes, including LDPC(49,24), LDPC(121,60), LDPC(121,80), Polar(32,16), Polar(64,32), Polar(128,86),  BCH(31,16), BCH(63,36), and BCH(127,120). We analyzed the rank distribution of the $\textbf{\textit{Q}} \cdot \textbf{\textit{K}}^T$ matrix across 1000 batches for the six-layer Transformer encoder. As illustrated in Figure~\ref{fig_ecct_rank}, the rank ranges from 12 to 181, with over 90\% of values below 69, consistently indicating a low-rank structure across all layers and code types. This observation suggests that the high-dimensional self-attention matrix can be approximated with a lower-rank representation, offering a pathway to reduce computational overhead.

\begin{figure}[htb]
\centering
\includegraphics[width=1.0\textwidth]{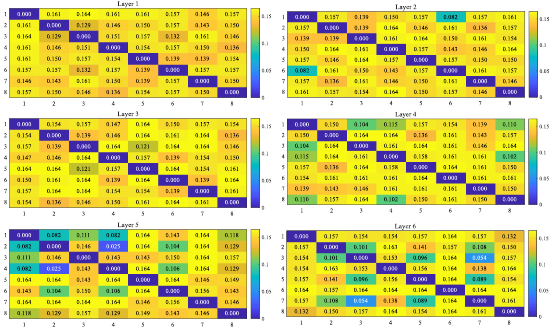}
\caption{Distribution of Jensen-Shannon divergence between attention head pairs across 6 layers, showing high similarity (JSD $<$ 0.17).}
\label{fig_jsd_sim}
\end{figure}

Furthermore, the conventional multi-head self-attention mechanism computes independent attention scores for each head using distinct weight matrices $\textbf{\textit{W}}_i^Q$, $\textbf{\textit{W}}_i^K$, and $\textbf{\textit{W}}_i^V$. However, we hypothesize that the attention distributions across heads may exhibit high similarity, particularly in channel decoding, where the sparse structure of the parity-check matrix $\mathbf{H}$ constrains bit interactions. To measure attention head similarity, we employed the Jensen-Shannon Divergence (JSD)~\cite{Jensen_Shannon_divergence}, a symmetric measure of similarity between probability distributions, defined as:
\begin{equation}
\label{eq_jsd}
\begin{aligned}
\text{JSD}(\mathbf{A}_i[k] \Vert \mathbf{A}_j[k]) = 
\frac{1}{2} \left[ D_{KL}(\mathbf{A}_i[k] \Vert \mathbf{M}[k]) + D_{KL}(\mathbf{A}_j[k] \Vert \mathbf{M}[k]) \right],
\end{aligned}
\end{equation}
where $k$ is the row index, $\mathbf{A}_i$ and $\mathbf{A}_j$ $\in \mathbb{R}^{N \times N}$are the softmax outputs of attention heads $i$ and $j$, $\mathbf{M}[k] = \frac{1}{2}(\mathbf{A}_i[k] + \mathbf{A}_j[k])$ is the average distribution, and $D_{KL}(\mathbf{P} \Vert \mathbf{Q}) = \sum_{l=1}^N \mathbf{P}[l] \log \frac{\mathbf{P}[l]}{\mathbf{Q}[l]}$ is the Kullback-Leibler divergence~\cite{Kullback_Leibler_divergence}. Using the same simulation setup as the low-rank validation ($L=6$, $H=8$, $d_k=182$, $d_f=2912$), we computed the per-row JSD for all head pairs across all rows, averaging the results. As shown in Figure~\ref{fig_jsd_sim}, JSD values are consistently below 0.17 across all layers, with a mean of 0.13, indicating strong alignment in attention distributions. This similarity, driven by the sparse structure of $\mathbf{H}$, supports the feasibility of sharing a common attention matrix across heads.

Motivated by these findings, we propose a novel \textbf{unified attention module (UAM)} to address the computational inefficiency and lack of cross-code generalization in traditional self-attention. The UAM introduces learnable memory components, $\textbf{\textit{A}}_l$ and $\textbf{\textit{V}}_l \in \mathbb{R}^{N \times d_l}$, where $d_l$ is a hyperparameter controlling the low-rank approximation. The UAM is defined 
as:
\begin{equation}
\label{eq_unified_attn}
\text{UniAttn}\left( {{\textbf{\textit{A}}_l},{\textbf{\textit{V}}_l}} \right) = \text{Softmax} \left( {{\textbf{\textit{A}}_l}} \right) \cdot \left( {{\textbf{\textit{V}}_l^T} \textbf{\textit{X}}} \right),
\end{equation}
where $\textbf{\textit{X}} \in \mathbb{R}^{N \times d_k}$ is the input to the attention layer. In contrast, the conventional self-attention is expressed as:
\begin{equation}
\label{eq_attn_1}
\begin{aligned}
\text{Attn} \left(\textbf{\textit{W}}^{\textit{Q}},\textbf{\textit{W}}^{\textit{K}},\textbf{\textit{W}}^{\textit{V}} \right)=
\text{Softmax} \left( \frac{\left(\textbf{\textit{XW}}^{\textit{Q}}\right) \cdot {\left(\textbf{\textit{XW}}^{\textit{K}}\right)}^T}{\sqrt{d_k}} \right) \cdot \left( \textbf{\textit{XW}}^{\textit{V}} \right).
\end{aligned}
\end{equation}

By eliminating the projections $\textbf{\textit{W}}^{\textit{Q}}$ and $\textbf{\textit{W}}^{\textit{K}}$, the UAM reduces the computational complexity from $\mathcal{O}(N^2 \cdot d_k + N \cdot d_k^2)$ to $\mathcal{O}(N \cdot d_l \cdot d_k)$, where $d_l \ll N$. The \textbf{shared attention matrix} $\mathbf{A}_l$ is applied across all heads, leveraging the observed head similarity to promote a unified decoding architecture. This design not only reduces the parameter count but also enables the model to learn shared features across diverse code types, facilitating a code-agnostic framework.

\begin{figure}[htb]
\centering
\includegraphics[width=0.6\columnwidth]{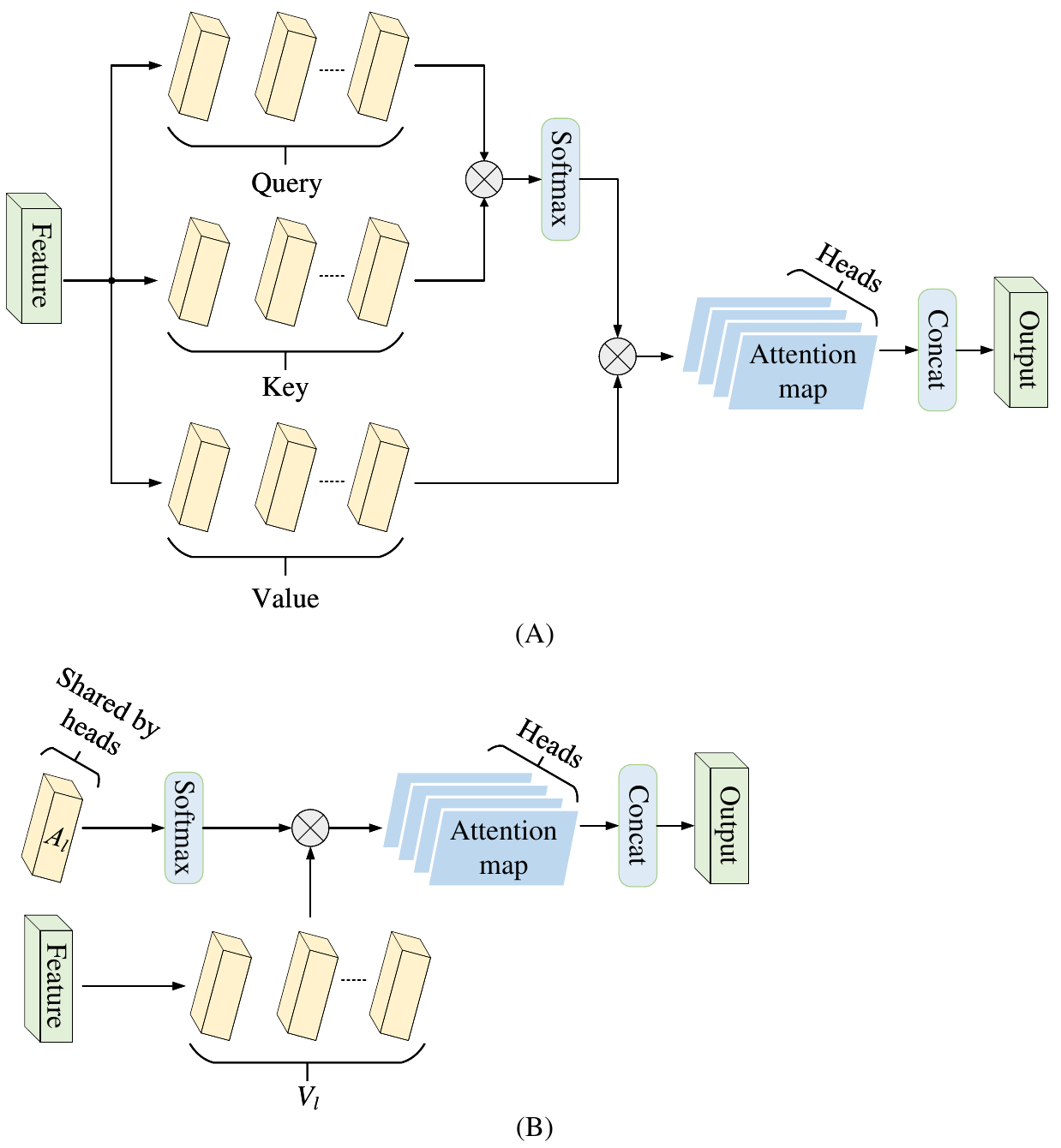}
\caption{Architecture of the vanilla multi-head self-attention (A) and the proposed multi-head unified-attention (B).}
\label{fig_arch_uam}
\end{figure}

\renewcommand{\arraystretch}{1.25}
\setcounter{magicrownumbers}{0} 
\begin{table}
\begin{center}
\label{Algorithm1}
\begin{tabular}{p{360pt}}
\toprule
\textbf{Algorithm 1} Pseudocode for Multi-Head Unified-Attention \\
\midrule
\rownumber \quad
Input: $\textbf{\textit{X}}$ \quad \# $\text{shape}=(B,N,d_h)$ \\
\rownumber \quad
Trainable Parameters: $\textbf{\textit{A}}_l$, $\textbf{\textit{V}}_l$ \\
\rownumber \quad
Hyper Parameters: number of heads $H$; batch size $B$ \\
\rownumber \quad
Hyper Parameters: embedding size $d_k$; $d_h = H \times d_k$ \\
\rownumber \quad
$\textbf{\textit{X}} = \textbf{\textit{X}}.\text{view}(B,N,H,d_k).\text{transpose}(1,2)$\\
\rownumber \quad
$\textbf{\textit{A}} = \text{Softmax}(\textbf{\textit{A}}_l, \text{dim}=-1)$ \quad
\# $\text{shape}=(B,1,N,d_l)$ \\
\rownumber \quad
$\textbf{\textit{X}}_{o} = \textbf{\textit{A}} \cdot \left( \textbf{\textit{V}}_l^T \cdot \textbf{\textit{X}} \right)$ \quad
\# $\text{shape}=(B,H,N,d_k)$ \\
\rownumber \quad
$\textbf{\textit{X}}_{o} = \textbf{\textit{X}}_{o}.\text{transpose}(1,2).\text{view}(B,N,d_h)$ \\
\rownumber \quad
$\textbf{\textit{X}}_{o} = \textbf{\textit{W}}^O(\textbf{\textit{X}}_{o})$ \\
\rownumber \quad
Output: $\textbf{\textit{X}}_{o}$ \quad \# $\text{shape}=(B,N,d_h)$ \\
\bottomrule
\end{tabular}
\end{center}
\end{table}
\setcounter{magicrownumbers}{0}

The architecture of the proposed multi-head UAM is depicted in Figure~\ref{fig_arch_uam}, with the corresponding pseudocode provided in Algorithm 1. The shared attention mechanism, combined with the low-rank approximation, significantly reduces model complexity while preserving decoding accuracy, as validated in subsequent experiments. This approach marks a significant advancement over the vanilla Transformer-based ECCT~\cite{ecct_2022}, which lacks cross-code unification.

\subsection{Sparse Masked Unified Attention}
Error detection and correction in channel decoding rely on analyzing received codewords against parity-check equations, where a non-zero syndrome indicates channel errors. In Transformer architectures, vanilla self-attention is inherently dense, correlating all bits indiscriminately, despite parity bits not interacting with every information bit. To address this, we propose a sparse masked unified attention mechanism that incorporates code-specific dependencies, inspired by human intelligence in discerning bit relationships. This approach accelerates the model's understanding of decoding principles, enhancing performance.

Specifically, for any linear block code defined by a parity-check matrix $\textbf{H} \in \mathbb{F}_2^{(n-k) \times n}$, we define a mapping $\text{M}(\textbf{H}): \{0,1\}^{(n-k) \times n} \to \{-\infty, 0\}^{N \times d_l}$ that generates a mask for the attention memory $\textbf{\textit{A}}_l$. The UAM is formulated as:
\begin{equation}
\label{eq_attn_mask}
\begin{aligned}
\text{UniAttn}\left(\boldsymbol{A}_l,\boldsymbol{V}_l\right)=
\text{Softmax}\left(\boldsymbol{A}_l+\text{M}(\textbf{H})\right) \cdot \left(\boldsymbol{V}_l^T\boldsymbol{X}\right).
\end{aligned}
\end{equation}

The design of the mask function $\text{M}(\textbf{H})$ aims to enable the neural network to capture the relationship between the received signal $\textbf{\textit{y}} \in \mathbb{R}^n$ and its syndrome $\text{s}(\textbf{\textit{y}}) = \textbf{H} \cdot \overline{\textbf{\textit{y}}} \in \{0,1\}^{n-k}$, where $\overline{\textbf{\textit{y}}} = \text{bin}(\text{sign}(\textbf{\textit{y}}))$ is the hard-decoded binary vector derived from $\textbf{\textit{y}}$. For a valid codeword $\textbf{\textit{y}}$ (i.e., noise-free or correctly decoded), the orthogonality condition $\textbf{H} \cdot \overline{\textbf{\textit{y}}} = \textbf{0}$ holds over $\mathbb{F}_2$, yielding $\text{s}(\textbf{\textit{y}}) = \textbf{0}$. To embed this relationship within the attention mechanism, we define an extended parity-check matrix $\overline{\textbf{H}} \in \mathbb{R}^{(2n-k) \times (n-k)}$ satisfying:
\begin{equation}
\label{eq_mask_cons}
\begin{aligned}
\left[\overline{\textbf{\textit{y}}}, \text{s}\left( \textbf{\textit{y}} \right) \right] \cdot \overline{\textbf{H}}=\textbf{0}.
\end{aligned}
\end{equation}

Inspired by this, we explore a mask structure leveraging the parity-check constraints. Let $\textbf{\textit{A}}_l$ be a matrix in $\mathbb{R}^{N \times d_l}$, where $N$ is defined as $2n-k$ and $d_l$ as $n-k$. The theoretical foundation for setting $d_l = n-k$ stems from the maximum rank of the parity-check matrix $\textbf{H} \in \mathbb{F}_2^{(n-k) \times n}$, which is $n-k$. We then construct the extended parity-check matrix $\overline{\textbf{H}}$ as follows:
\begin{equation}
\label{eq_gen_mask}
\begin{aligned}
\overline{\textbf{H}}=\begin{bmatrix}{\textbf{H}}^T\\{\textbf{\textit{I}}}_{n-k}\end{bmatrix},
\end{aligned}
\end{equation}
where ${\textbf{\textit{I}}}_{n-k} \in \mathbb{R}^{(n-k) \times (n-k)}$ is an identity matrix with ones on the diagonal and zeros elsewhere. The generation of the attention mask is thus a direct process, formulated as $\text{M}(\overline{\textbf{H}})$.

We naturally proceed to define the density of $\overline{\textbf{H}}$ as:
\begin{equation}
\label{eq_h_density}
\begin{aligned}
H_d=\frac{\text{sum}(\overline{\textbf{H}})}{(2n-k) \times (n-k)}.
\end{aligned}
\end{equation}

\begin{figure}[htb]
\centering
\includegraphics[width=0.7\columnwidth]{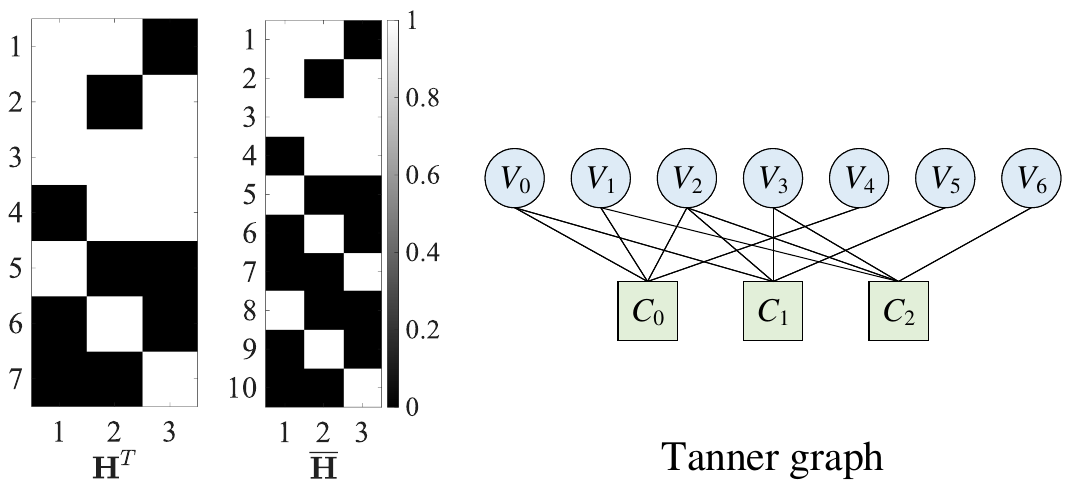}
\caption{We present the corresponding matrices $\textbf{H}^T$, $\overline{\textbf{H}}$ and the Tanner graph for the Hamming (7,4) code.}
\label{fig_hamming}
\end{figure}

For enhanced clarity, Figure \ref{fig_hamming} offers a visual depiction of matrices $\textbf{H}^T$, $\overline{\textbf{H}}$ and the Tanner graph, utilizing the Hamming (7,4) code as an illustrative example. Here, matrix $\textbf{H}$ is defined as:
\begin{equation}
\label{eq_h_matrix}
\begin{aligned}
\textbf{H} = \begin{bmatrix}
1 & 1 & 1 & 0 & 1 & 0 & 0 \\
1 & 0 & 1 & 1 & 0 & 1 & 0 \\
0 & 1 & 1 & 1 & 0 & 0 & 1
\end{bmatrix}.
\end{aligned}
\end{equation}

\renewcommand{\arraystretch}{1.25}
\setcounter{magicrownumbers}{0} 
\begin{table}
\begin{center}
\label{Algorithm2}
\begin{tabular}{p{360pt}}
\toprule
\textbf{Algorithm 2} Pseudocode for Sparse Mask Construction\\
\midrule
\rownumber \quad
Input: $\textbf{H}$ \quad \# $\text{shape}=(n-k,n)$ \\
\rownumber \quad
$\overline{\textbf{H}} = \text{zeros}(2n-k,n-k)$ \\
\rownumber \quad
$\overline{\textbf{H}}[0:n, 0:n-k] = \textbf{H}.\text{transpose}(0,1)$ \\
\rownumber \quad
$\overline{\textbf{H}}[n:2n-k, 0:n-k] = \text{eye}(n-k)$ \\
\rownumber \quad
Output: $(-\infty) \cdot (\neg\mathrm{~\overline{\textbf{H}}})$ \\
\bottomrule
\end{tabular}
\end{center}
\end{table}
\setcounter{magicrownumbers}{0}

This illustration demonstrates that the application of the mask results in sparser attention, a phenomenon particularly pronounced in low-density parity-check codes. Most importantly, the computational complexity of the UAM is now reduced to the density of the code $\mathcal{O}(N \times d_l \times d_k \times H_d)$. Our experimental results indicate that the use of this mask has successfully reduced the computational complexity of the attention mechanism by an average of $\textbf{86\%}$. Furthermore, we provide a summary of the Python-style sparse mask construction within Algorithm 2. It should be noted that, in application-specific integrated circuit (ASIC) implementations, computations involving the masked parts can be directly \textbf{\textit{skipped}} to enhance efficiency~\cite{hardware_acc}.

\subsection{Architecture and Training Methodology}
\label{subsec:arch_train_method}
\begin{figure}[htb]
\centering
\includegraphics[width=1.0\textwidth]{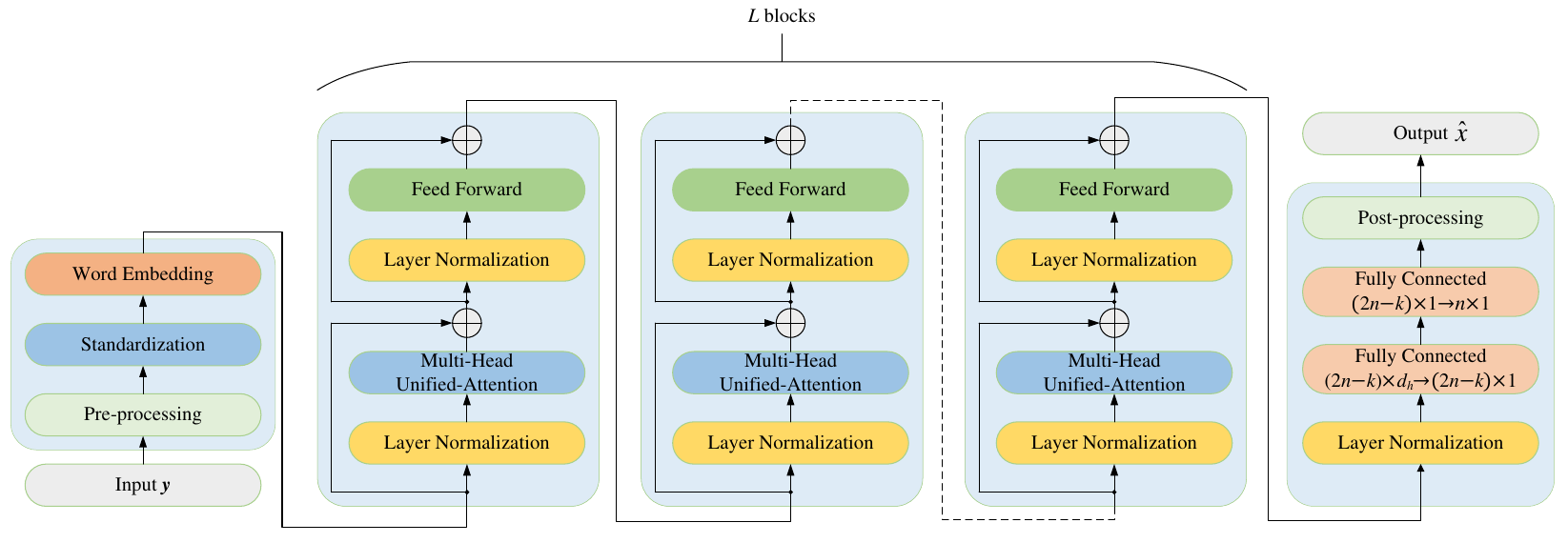}
\caption{Illustration of the proposed unified error correction code Transformer.}
\label{fig_arch_full}
\end{figure}

The architecture of the proposed unified error correction code Transformer is illustrated in Figure \ref{fig_arch_full}. The primary innovation in our model involves substituting the vanilla attention mechanism with a multi-head shared unified-attention module. This modification not only decreases computational complexity but also empowers the model to identify distinctive features among diverse codewords, thereby facilitating the establishment of a unified decoding architecture. We employ fine-tuning techniques to enhance the model’s generalization to unseen codewords. Furthermore, we have designed standardization units to standardize the lengths of codewords and syndromes across diverse datasets, ensuring the model's compatibility with various code types, lengths, and rates.

In Figure \ref{fig_arch_full}, the output $\textbf{\textit{y}}$ from the AWGN channel passes through the pre-processing module to yield $N=2n-k$ input elements for the neural network. The decoder is constructed by concatenating $L$ Transformer encoders, which consist of unified-attention and feed-forward layers, with normalization layers in between. The output module is characterized by two fully connected layers. The initial layer compresses the multi-head output into a one-dimensional vector of size $2n-k$, with the subsequent layer transforming it into an $n$-dimensional vector representing the softly decoded noise. Finally, the post-processing module acts on the estimated noise to derive the transmitted original codeword.

We employ binary cross-entropy as the loss function with the objective of learning to predict the multiplicative noise $\textbf{\textit{z}}$. The corresponding target binary multiplicative noise is denoted as $\tilde {\textbf{\textit{z}}}=\text{bin}(\text{sign}(\textbf{\textit{z}}))$. Therefore, the loss calculation for a received individual codeword $\textbf{\textit{y}}$ is given by:
\begin{equation}
\label{eq_loss}
\begin{aligned}
\text{loss} = -\sum_{i=1}^{n} {\tilde{z}}_i \text{log}(f_{\theta}(y)) + (1-{\tilde{z}}_i) \text{log}(1 - f_{\theta}(y)).
\end{aligned}
\end{equation}

To accommodate codewords of varying lengths within a unified architecture, we have designed a standardization unit that performs padding on $\lvert \textbf{\textit{y}} \rvert$ and $s(\textbf{\textit{y}})$ as described in Equation (\ref{eq_pre_process}). Specifically, after receiving the channel output, we pad different codewords and their corresponding syndromes with zeros up to the maximum length. Assuming that the maximum length of the codeword is $N_{\text{max}}$ and the maximum length of the syndrome is $S_{\text{max}}$, the dataset is standardized according to:
\begin{equation}
\label{eq_padding}
\begin{aligned}
\widetilde {\textbf{\textit{y}}} = \left [~ \left [\lvert \textbf{\textit{y}} \rvert,~\textbf{0}_c \right ],~ \left [s(\textbf{\textit{y}}),~\textbf{0}_s \right ]~ \right],
\end{aligned}
\end{equation}
where $\textbf{0}_c$ and $\textbf{0}_s$ represent zero vectors with sizes $N_{\text{max}}-\text{size}(\textbf{\textit{y}})$ and $S_{\text{max}}-\text{size}(s(\textbf{\textit{y}}))$, respectively. Through this unit, the Transformer neural decoder can accommodate any codeword within the maximum length, allowing for joint training in a unified manner. In ASIC implementations, computations for zero-padded parts can be skipped to improve efficiency, similar to the sparse-masked attention~\cite{hardware_acc}.

In the training phase, we conducted 1000 epochs, each consisting of 1000 minibatches, which in turn contained 512 samples each. Samples for each minibatch were randomly selected from a pool of LDPC, Polar, and BCH codewords. The learning rate was initialized at $10^{-3}$, and a cosine decay scheduler was applied without warmup~\cite{ml_research}, gradually reducing it to $10^{-6}$ by the end of the training process. We employ the Adam optimizer to dynamically adjust the learning rate, leveraging its efficacy in managing diverse data types and model architectures~\cite{stochastic_opt}. The AWGN noise introduced to the links had its signal-to-noise ratio (SNR) randomly chosen between 3 and 7 for each batch. All experiments were conducted on NVIDIA GeForce RTX 4090 24GB GPUs.

\section{Experiments}
We conducted experiments on linear block codes, specifically LDPC, Polar, and BCH, to assess the effectiveness of the proposed architecture. The parity-check matrices utilized in the paper were sourced from~\cite{database_channel_codes}. We benchmark our approach against the most recent state-of-the-art methods from~\cite{ecct_2022}, which trains each codeword parameter independently, and the jointly trained FECCT~\cite{fecct_2024}. Additionally, we compare it to legacy belief propagation (BP)-based decoders, including unlearned BP~\cite{cmp_bp}, fully supervised Hyper BP~\cite{cmp_hyper_bp}, and Autoregressive BP (AR BP)~\cite{cmp_arbp}. The results are presented in the form of bit error rate (BER) and block error rate (BLER) at various normalized SNR (i.e. $\mathrm{E_b/N_0}$) values, with at least $10^5$ random codewords tested per SNR. We define aggregate BER (ABER) and aggregate BLER (ABLER) as the overall error rates spanning all evaluated codewords.

In addition, to enable direct comparison with ECCT and FECCT, whose simulation results were sourced from their respective papers~\cite{ecct_2022, fecct_2024}, we fine-tuned the pre-trained UECCT models on consultative committee for space data systems (CCSDS)~\cite{ccsds} and MacKay~\cite{mackay} codewords using jointly trained LDPC, Polar, and BCH models. Given GPU memory constraints and the fact that the performance of long codes has been thoroughly explored~\cite{ldpc_capacity}, while the error correction of short codes remains far from the Shannon limit~\cite{shannon_limit}, this work primarily focuses on training and validating \textbf{\textit{medium-to-short codes}}.

The proposed unified error correction code Transformer is defined by the following hyperparameters: the number of Transformer encoder layers $L$, the number of unified attention heads $H$, the dimensions of word embeddings $d_k$, the dimensions $d_l$ of the trainable memory $\textbf{\textit{A}}_l$ and $\textbf{\textit{V}}_l$, and the dimension $d_f$ of the FFN layer. We present the performance of our framework with hyperparameters are set to $L = 6$, $H = 8$, $d_k = 64$, $d_l = 64$, and $d_f = 4 \cdot H \cdot d_k$. The value of $d_l = 64$ corresponds to the maximum syndrome length among the jointly trained codewords.

\begin{table}[htb]
\centering
\caption{Negative natural logarithm of BER for ECCT, FECCT, and UECCT across Polar, LDPC, BCH, CCSDS, and MacKay codes at $\mathrm{E_b/N_0}$ = 4, 5, 6 dB, with best results in bold (higher is better).}
\includegraphics[width=1.0\textwidth]{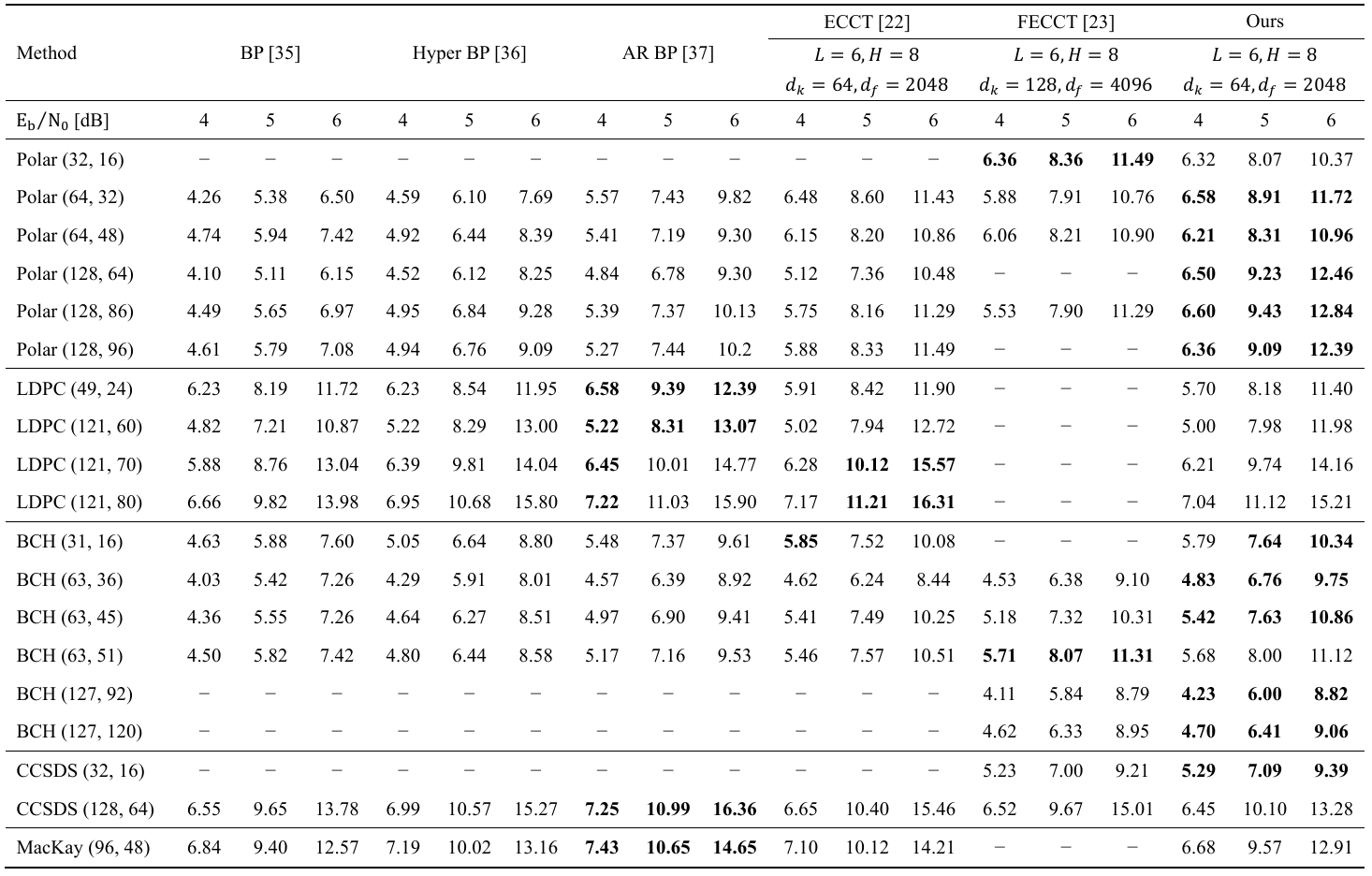}
\label{tab_sim_result}
\end{table}

\begin{figure}[htb]
\centering
\includegraphics[width=0.8\columnwidth]{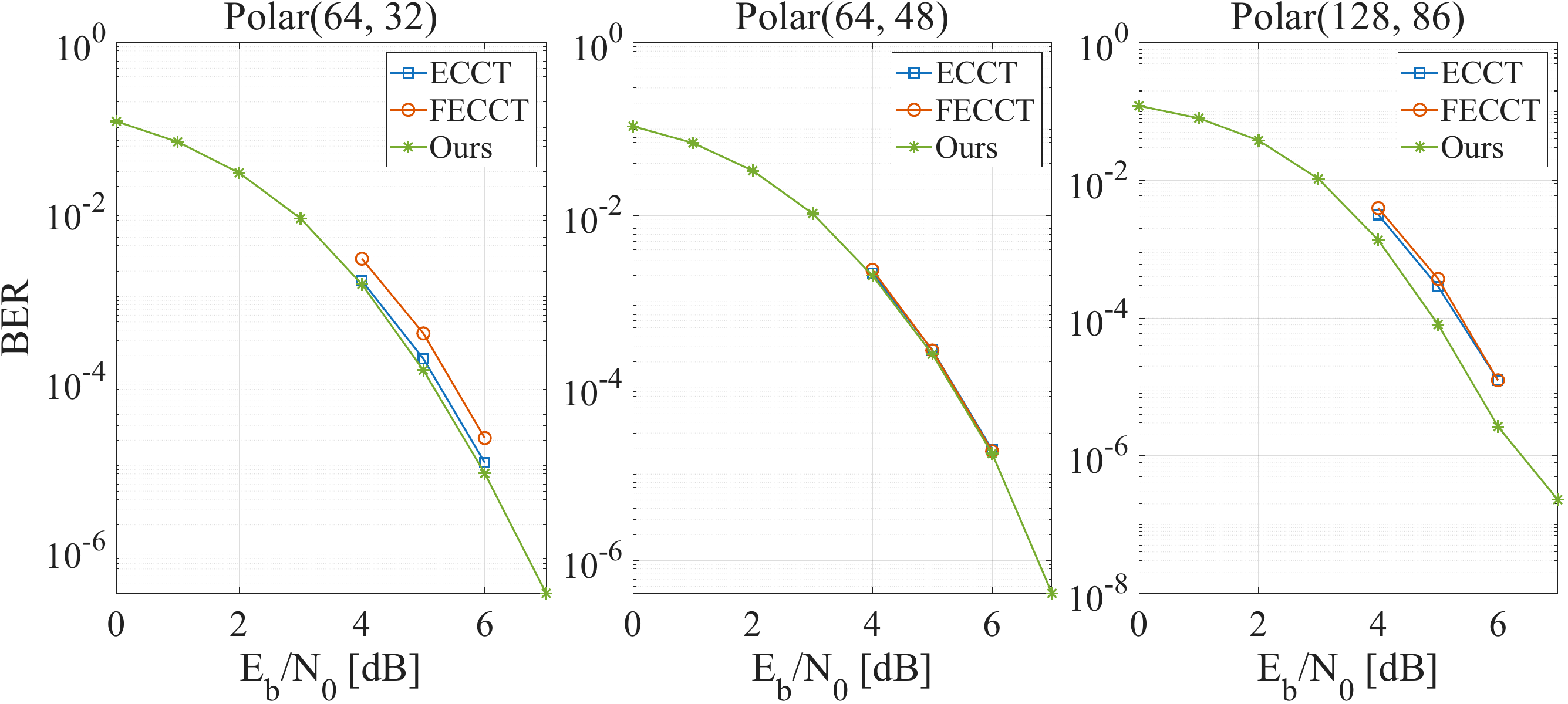}
\caption{BER comparison of ECCT, FECCT, and proposed UECCT for Polar codes.}
\label{fig_polar_sim_result}
\end{figure}

\begin{figure}[htb]
\centering
\includegraphics[width=0.8\columnwidth]{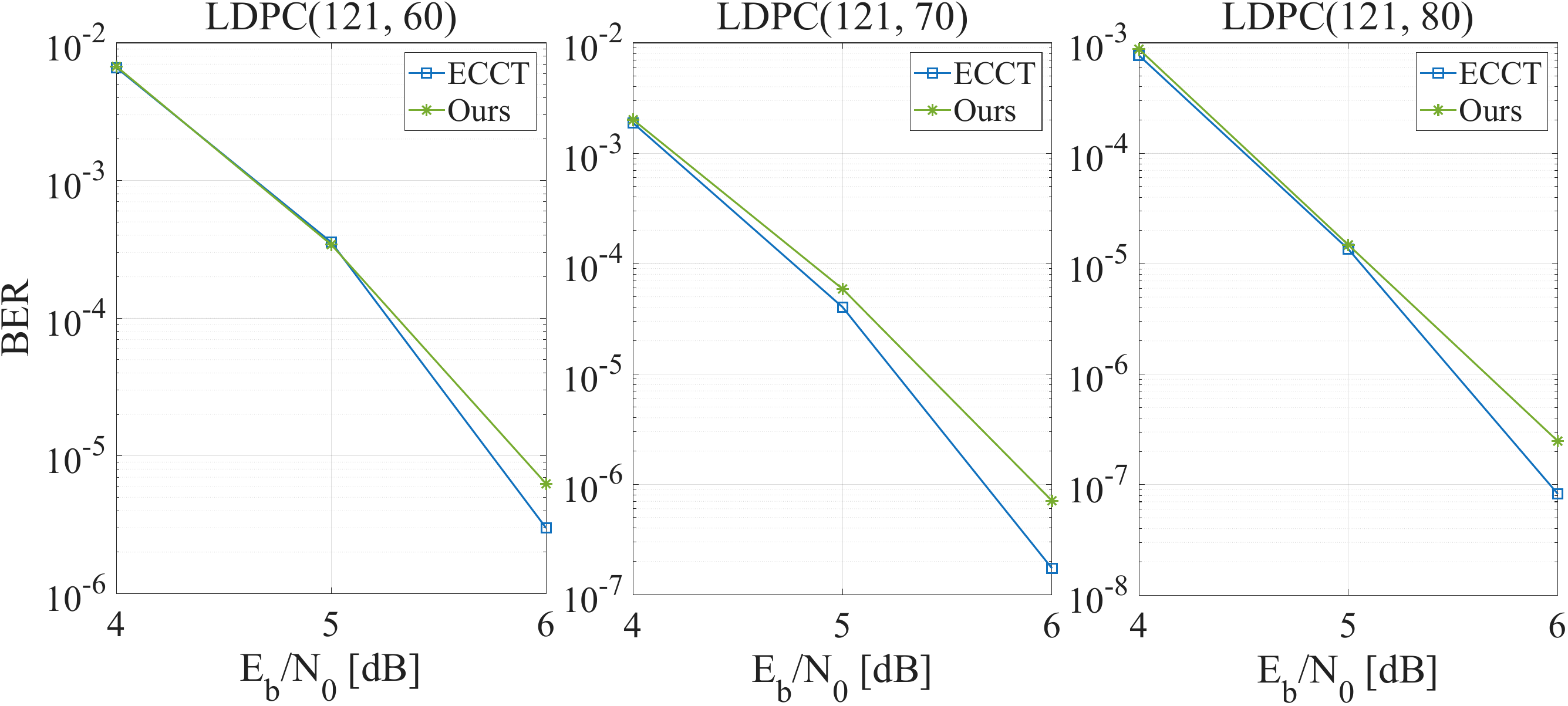}
\caption{BER comparison of ECCT and proposed UECCT for LDPC codes.}
\label{fig_ldpc_sim_result}
\end{figure}

\begin{figure}[htb]
\centering
\includegraphics[width=0.8\columnwidth]{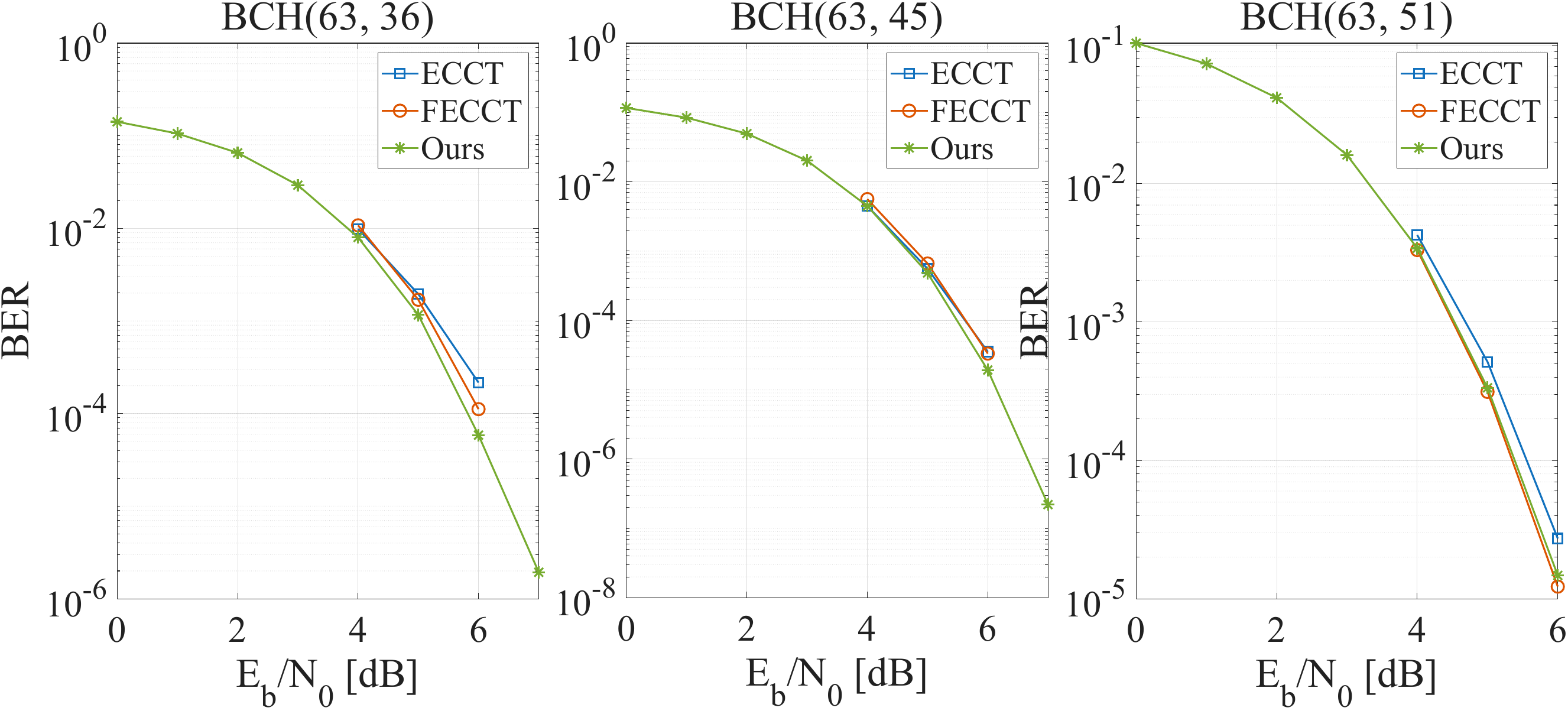}
\caption{BER comparison of ECCT, FECCT, and proposed UECCT for BCH codes.}
\label{fig_bch_sim_result}
\end{figure}

Table \ref{tab_sim_result} presents the simulation results, specifically showing the negative natural logarithm of the BER at $\mathrm{E_b/N_0}$ values of 4, 5, and 6 dB. It is noted that higher values in this context indicate better performance, with the best results for each SNR highlighted in bold. Additional BER plots for Polar, LDPC and BCH codes are provided in Figure \ref{fig_polar_sim_result}, \ref{fig_ldpc_sim_result} and Figure \ref{fig_bch_sim_result}, respectively. Simulation results demonstrate that our proposed UECCT outperforms both the individually trained ECCT~\cite{ecct_2022} and the jointly trained FECCT~\cite{fecct_2024} across the majority of codewords, validating the effectiveness of the proposed architecture. Notably, UECCT even surpasses FECCT’s performance with an embedding length of 128 when operating at a more compact embedding length of 64, further highlighting the computational efficiency of our design. The superior performance arises from its ability to utilize a shared attention mechanism across heads, combined with sparsity-aware masking that reduces computational complexity while maintaining decoding accuracy.

\begin{figure}[htb]
\centering
\includegraphics[width=0.8\columnwidth]{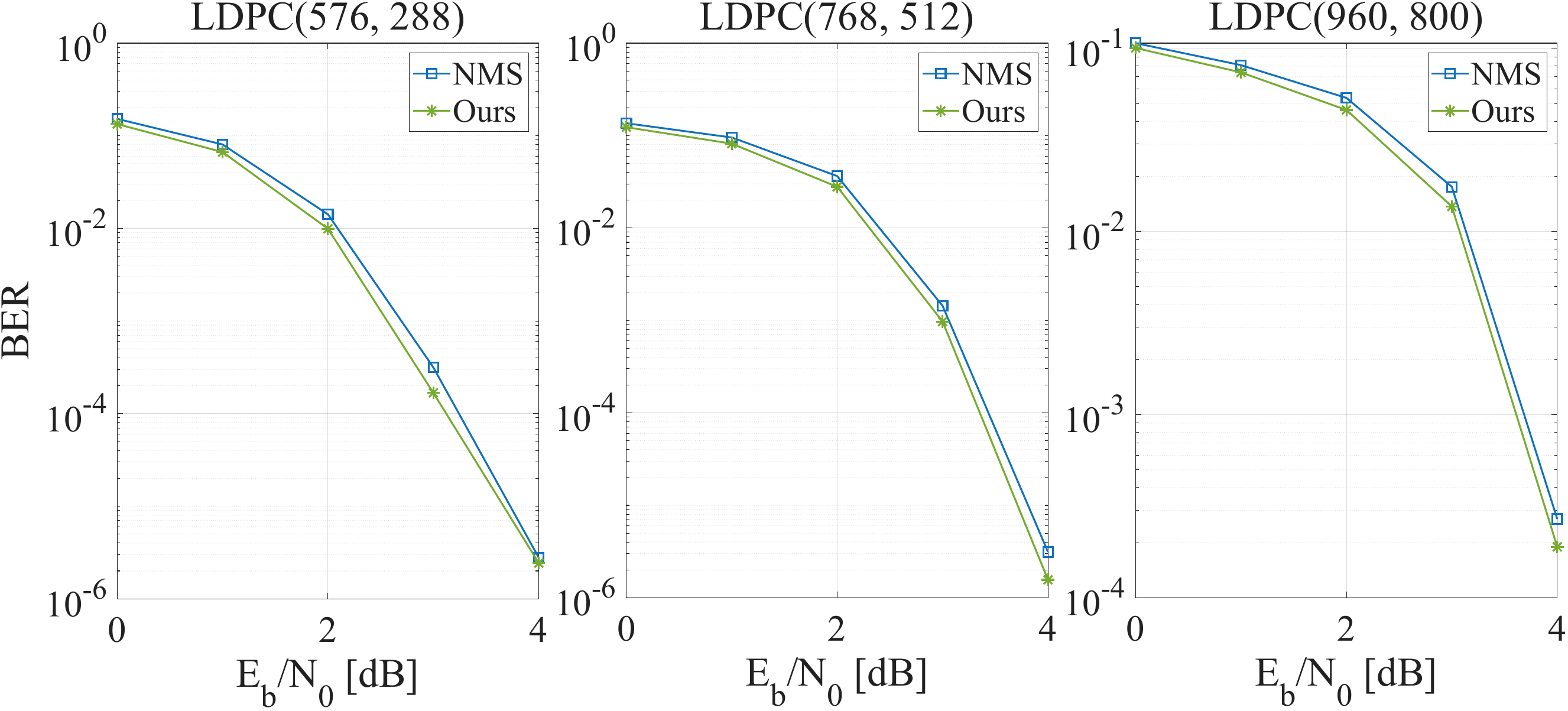}
\caption{BER comparison between the NMS algorithm and the proposed UECCT for medium-to-long LDPC codes.}
\label{fig_nms_sim_result}
\end{figure}

To further validate the performance and scalability of the proposed method on medium-to-long codes, we conducted joint training using LDPC(576, 288), LDPC(768, 512), and LDPC(960, 800) codewords with hyperparameters set to $L=12$, $H=8$, $d_k=16$, $d_l=288$, and $d_f=512$. The neural network training and inference were performed under an AWGN channel with BPSK modulation. During training, the SNR was randomly selected between 2 dB and 3 dB, with a batch size of 192, while other training methods were consistent with those described in Section \ref{subsec:arch_train_method}. During inference, $10^5$ code blocks were tested per SNR. For comparison with traditional decoding algorithms, we utilized the normalized min-sum (NMS) algorithm for LDPC codes~\cite{ldpc_nms}. The NMS algorithm was configured with a normalization factor of 0.875 and a maximum of 15 iterations. Additionally, an early stopping mechanism was implemented to halt decoding once the parity-check matrix $\mathbf{H}$ is satisfied. Figure~\ref{fig_nms_sim_result} presents the BER performance comparison between the NMS algorithm and the proposed UECCT. The simulation results demonstrate that the proposed method achieves performance comparable to the traditional NMS decoding algorithm. This indicates that our decoding architecture effectively scales to medium-to-long codes.

\section{Analysis}
Using the parameters in Table \ref{tab_sim_result}, we examine the effect of Transformer encoder layers on performance, focusing on the embedding dimension $d_k$ and sparse masked attention efficiency. We also compare our approach’s computational complexity with that of existing methods.

\subsection{Impact of Encoder Layers and Embedding Dimensions}
\begin{figure}[htb]
\centering
\includegraphics[width=0.6\columnwidth]{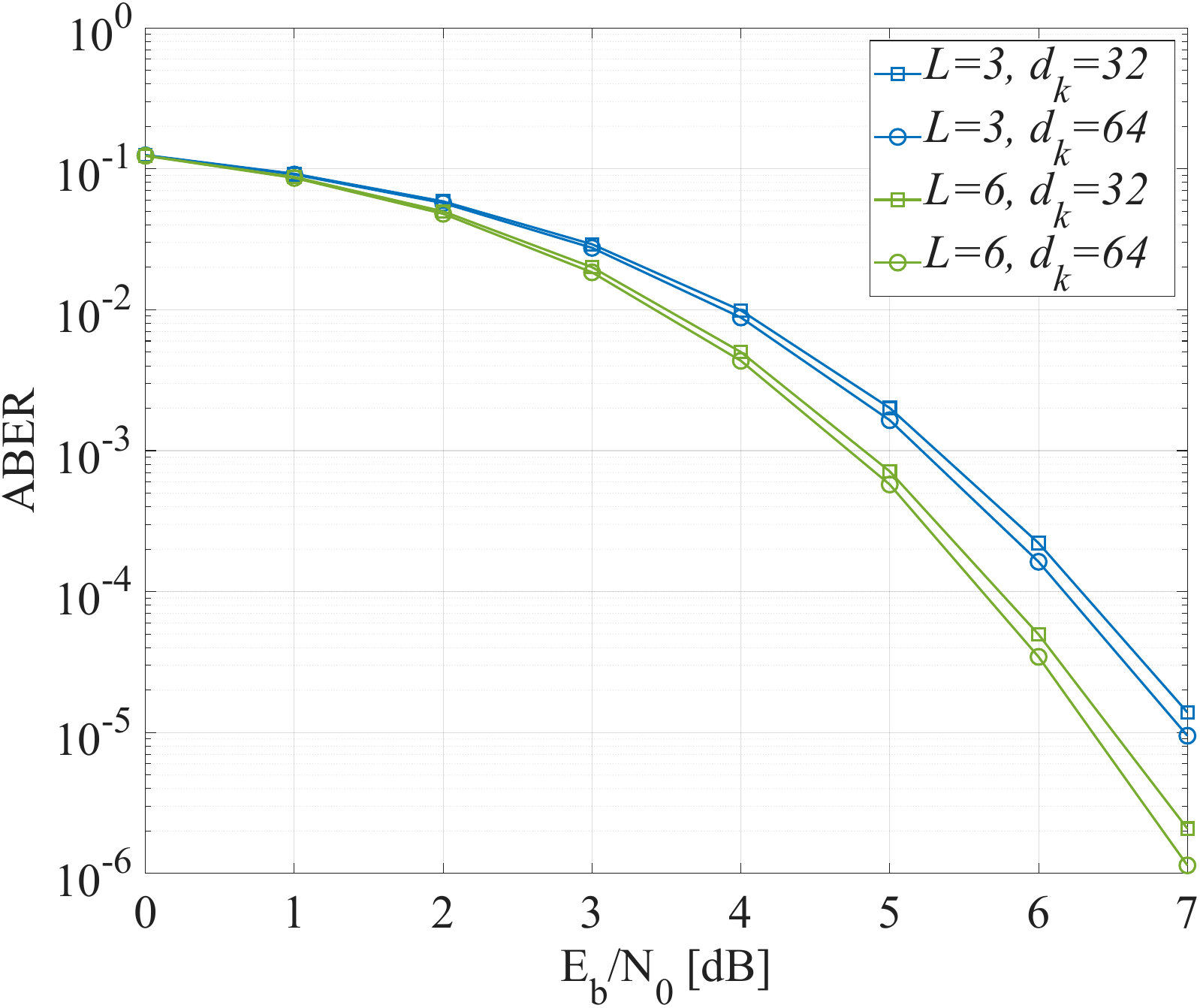}
\caption{ABER performance across various Transformer encoder layers and the embedding dimensions $d_k$.}
\label{fig_layer_sim_result}
\end{figure}

Figure \ref{fig_layer_sim_result} displays the ABER performance across varying encoder layers $L$ and embedding dimensions $d_k$. Results reveal that $L$ exerts a stronger influence on performance than $d_k$. Increasing $L$ broadens the parameter space, enhancing nonlinear fitting and enabling complex representations.

Deeper architectures better capture intricate data patterns and dependencies, crucial for error correction. Additional layers enable multi-level abstraction and refinement, improving error estimation and correction, especially in complex wireless communication systems where simpler models falter.

These findings highlight the primacy of model depth in Transformer-based error correction. Prioritizing $L$ over $d_k$ can optimize model design, enhancing practical performance.

\subsection{Impact of Sparse Masked Attention}
\begin{figure}
\centering
\includegraphics[width=0.9\columnwidth]{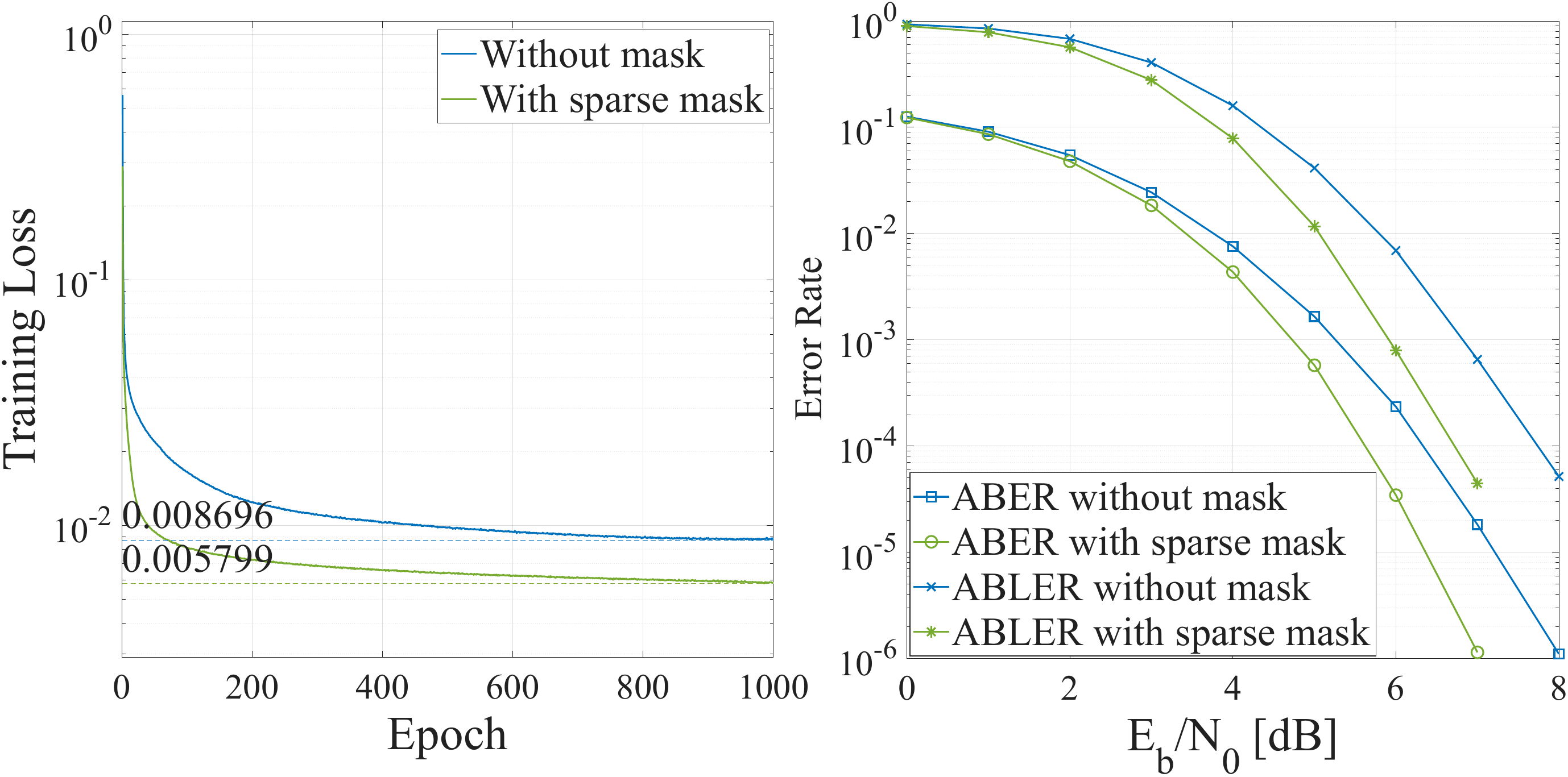}
\caption{The overall improvement in loss, ABER and ABLER with sparse masked attention.}
\label{fig_mask_sim_result}
\end{figure}

Figure \ref{fig_mask_sim_result} illustrates the effect of sparse masked attention on training loss, ABER, and ABLER. Sparse masked attention reduces training loss by 33.3\% and speeds up convergence. It also significantly boosts error correction, improving ABER by 0.7 dB at a bit error rate of $10^{-5}$ and ABLER by 0.9 dB at a block error rate of $10^{-4}$.

These gains highlight the effectiveness of embedding domain-specific knowledge into the attention mechanism, yielding significant improvements in efficiency and accuracy. By reducing computational complexity while boosting the model’s capacity to decode complex signals, this approach emerges as a promising advancement for future error correction techniques.

\subsection{Computational Complexity Analysis}
Before delving into the computational complexity, let's first review the unified error correction code Transformer presented in Figure \ref{fig_arch_full}, which  incorporates $L$ layers of Transformer encoders. The principal computational complexity within each layer is found in the multi-head self-attention module, which is predominantly responsible for matrix multiplication and is defined by the hyperparameters $H$, $d_k$ and $d_l$. Consequently, the computational complexity of the proposed architecture is given by $\mathcal{O}\left( L \times H \times \left( N \times d_l \times d_k \times H_d \right) \right)$, where $N=2n-k$, and $d_l \ll N$. Thanks to the sparsity of the parity-check matrix, $H_d$ is usually quite small. In our experiment, $H_d$ is approximately 0.14, which significantly reduces the computational complexity of the neural network. By contrast, the vanilla Transformer architectures in ECCT and FECCT have complexities of $\mathcal{O}\left( L \times H \times \left( N^2 \times d_k \times M_d + N \times d_k^2 \right) \right)$ and $\mathcal{O}\left( L \times H \times \left( N^2 \times d_k + N \times d_k^2 + N^2 \times M_d \right) \right)$, respectively, where $M_d$ represents the mask density in these models.

\begin{table}[htb]
\centering
\caption{Comparison of trainable parameters, MACs, training time, inference time, and mask density among ECCT, FECCT, and UECCT for various codes, all with identical parameters $L=6$, $H=8$, $d_k=64$, $d_l=64$, and $d_f=2048$.}
\includegraphics[width=1.0\textwidth]{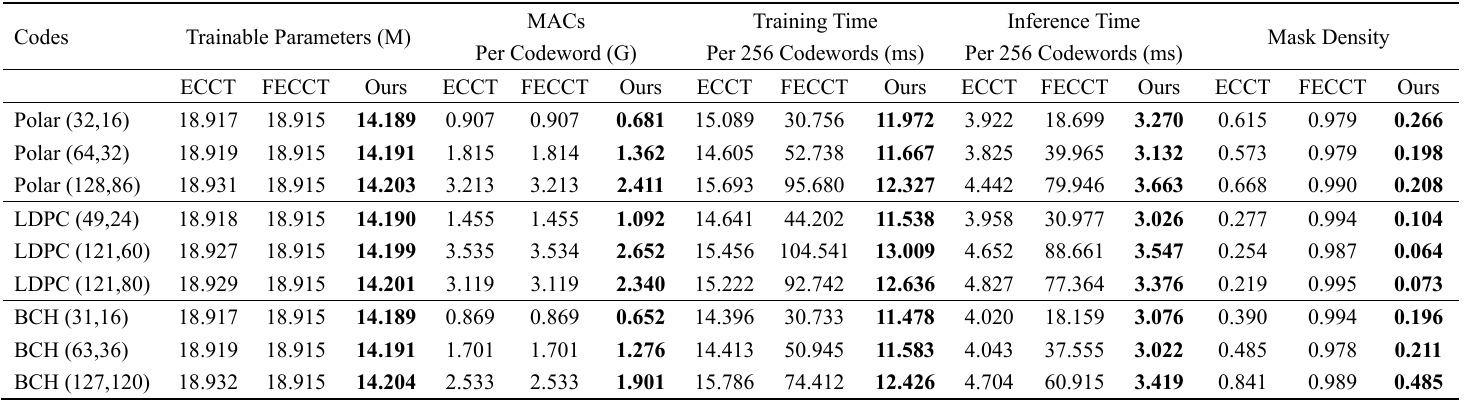}
\label{tab_para_flops}
\end{table}

\begin{figure}[htb]
\centering
\includegraphics[width=0.6\columnwidth]{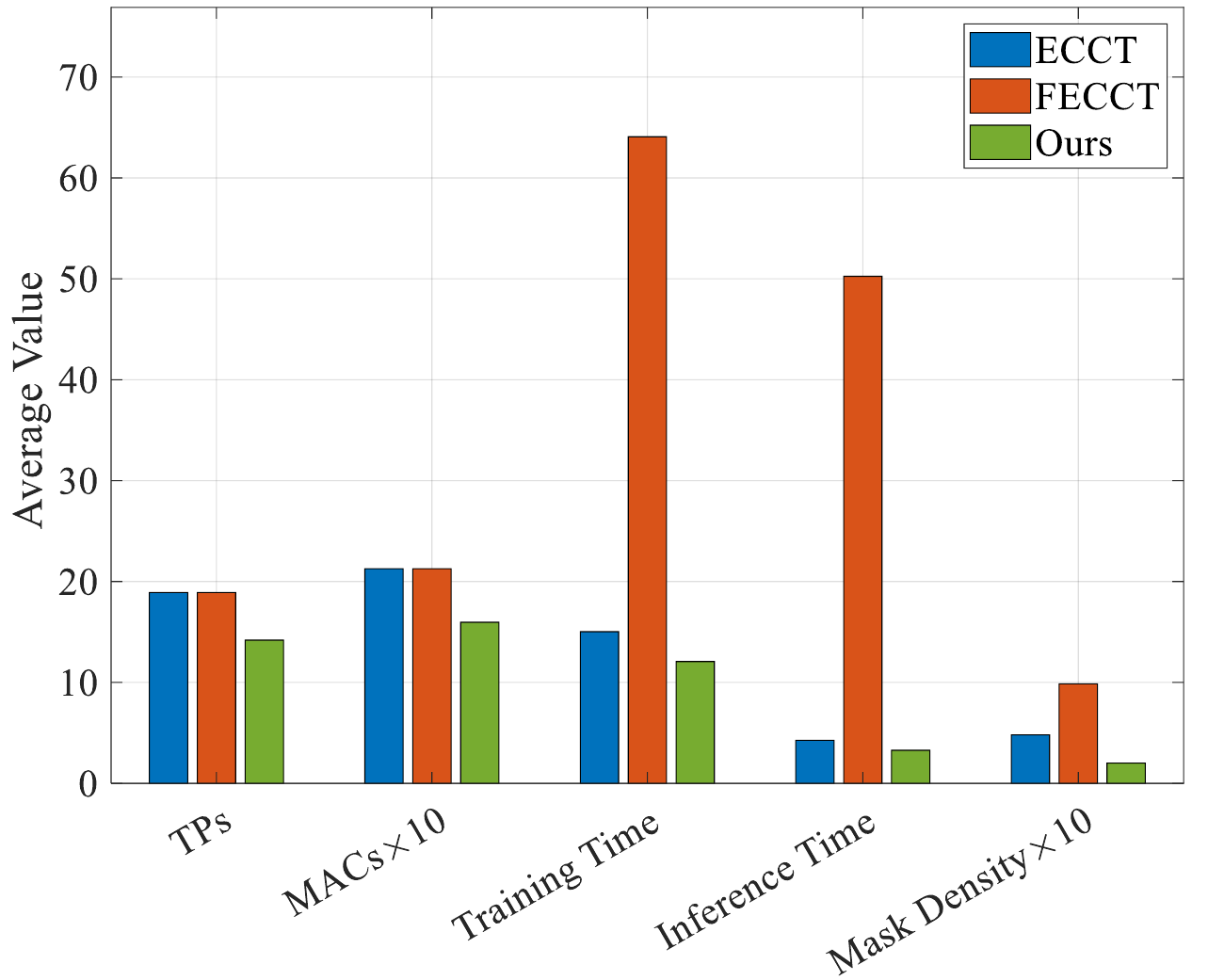}
\caption{Comparison of average TPs, MACs, training time, inference time, and mask density between ECCT, FECCT and UECCT.}
\label{fig_para_flops}
\end{figure}

Table \ref{tab_para_flops} compares the trainable parameters (TPs), multiply-accumulate operations (MACs), training time, inference time, and mask density for ECCT, FECCT, and UECCT. Lower values are preferable, with the best performing values highlighted in bold. The metrics for TPs and MACs are obtained using Thop~\cite{flops_thop}, a neural network profiling tool. UECCT reduces parameters and MACs by 25.0\% compared to both ECCT and FECCT through low-rank approximation, shared attention, and omission of $\textbf{\textit{Q}}$ and $\textbf{\textit{K}}$ projections. This results in training times 19.7\% and 81.2\% lower, and inference times 23.1\% and 93.5\% lower than ECCT and FECCT, respectively. Mask densities are influenced by design choices: FECCT has the highest density due to Tanner graph distance calculations, ECCT has a moderate density from extended bit-pair relations, and UECCT has the lowest density, 58.2\% and 79.7\% lower than ECCT and FECCT, achieved through the direct utilization of $\textbf{H}$ and the extension of the identity matrix. For a more intuitive representation, the average values of each metric across the codewords in Table \ref{tab_para_flops} are provided in Figure \ref{fig_para_flops}.

Compared to the approach in~\cite{ecct_2022, fecct_2024}, while the proposed method excels in decoding versatility and computational complexity, it may fall short in terms of memory efficiency, power consumption, and computational resource utilization when compared to non-learning solutions. This inefficiency could limit its potential for deployment. To address these issues, future implementations could leverage techniques such as parameter sharing, low-rank factorization, and quantization, aiming to reduce complexity and enhance efficiency~\cite{transformer_papa_share, low_rank_attn, transformer_quantization}.

\section{Conclusions}
We proposed a unified Transformer-based decoder that seamlessly integrates multiple linear block codes within a single framework. By introducing standardized unit, a unified attention module, and a sparse mask leveraging the parity-check matrix's sparsity, we achieved enhanced decoding accuracy and reduced computational complexity from $\mathcal{O}(N^2)$ to $\mathcal{O}(N)$. This work delivers a high-performance, low-complexity solution, addressing hardware overhead and scalability challenges in next-generation wireless systems like 6G. Future work will focus on improving memory efficiency and power consumption through techniques such as pruning and quantization to enhance deployability in resource-constrained environments.


\bibliographystyle{IEEEtran}
\bibliography{references}

\begin{thebibliography}{10}
\providecommand{\url}[1]{#1}
\csname url@samestyle\endcsname
\providecommand{\newblock}{\relax}
\providecommand{\bibinfo}[2]{#2}
\providecommand{\BIBentrySTDinterwordspacing}{\spaceskip=0pt\relax}
\providecommand{\BIBentryALTinterwordstretchfactor}{4}
\providecommand{\BIBentryALTinterwordspacing}{\spaceskip=\fontdimen2\font plus
\BIBentryALTinterwordstretchfactor\fontdimen3\font minus \fontdimen4\font\relax}
\providecommand{\BIBforeignlanguage}[2]{{%
\expandafter\ifx\csname l@#1\endcsname\relax
\typeout{** WARNING: IEEEtran.bst: No hyphenation pattern has been}%
\typeout{** loaded for the language `#1'. Using the pattern for}%
\typeout{** the default language instead.}%
\else
\language=\csname l@#1\endcsname
\fi
#2}}
\providecommand{\BIBdecl}{\relax}
\BIBdecl

\bibitem{massive_mimo_survey}
E.~Björnson, L.~Sanguinetti, H.~Wymeersch, J.~Hoydis, and T.~L. Marzetta, ``Massive {MIMO} is a reality—{What} is next? {Five} promising research directions for antenna arrays,'' \emph{Digit. Signal Process.}, vol.~94, pp. 3--20, Nov. 2019.

\bibitem{toward_6g}
M.~Giordani, M.~Polese, M.~Mezzavilla, S.~Rangan, and M.~Zorzi, ``Toward {6G} networks: Use cases and technologies,'' \emph{IEEE Commun. Mag.}, vol.~58, no.~3, pp. 55--61, Mar. 2020.

\bibitem{road_to_6g}
W.~Jiang, B.~Han, M.~A. Habibi, and H.~D. Schotten, ``The road towards {6G}: A comprehensive survey,'' \emph{IEEE Open J. Commun. Soc.}, vol.~2, pp. 334--366, Feb. 2021.

\bibitem{channel_coding_6g}
H.~Zhang and W.~Tong, ``Channel coding for {6G} extreme connectivity—requirements, capabilities, and fundamental tradeoffs,'' \emph{IEEE BITS Inf. Theory Mag.}, vol.~3, no.~1, pp. 1--12, Mar. 2024.

\bibitem{gallager_ldpc}
R.~Gallager, ``Low-density parity-check codes,'' \emph{IEEE Trans. Inf. Theory}, vol.~8, no.~1, pp. 21--28, Jan. 1962.

\bibitem{arikan_polar_code}
E.~Arikan, ``Channel polarization: A method for constructing capacity-achieving codes for symmetric binary-input memoryless channels,'' \emph{IEEE Trans. Inf. Theory}, vol.~55, no.~7, pp. 3051--3073, Jul. 2009.

\bibitem{ran1_coding_std}
3GPP, ``3gpp {TSG} {RAN} {WG1} meeting \#122-bis,'' 3rd Generation Partnership Project (3GPP), Technical Report, Oct. 2025.

\bibitem{bch_code}
R.~C. Bose and D.~K. Ray-Chaudhuri, ``On a class of error correcting binary group codes,'' \emph{Inf. Control}, vol.~3, no.~1, pp. 68--79, Mar. 1960.

\bibitem{3gpp_38212_std}
3GPP, ``Technical specification group radio access network; multiplexing and channel coding,'' 3rd Generation Partnership Project (3GPP), Technical Specification (TS) 38.212, Sep. 2023, 3GPP TS 38.212, V18.0.0.

\bibitem{hardware_ldpc}
M.~P.~C. Fossorier, M.~Mihaljevic, and H.~Imai, ``Reduced complexity iterative decoding of low-density parity check codes based on belief propagation,'' \emph{IEEE Trans. Commun.}, vol.~47, no.~5, pp. 673--680, May 1999.

\bibitem{hardware_sc}
C.~Leroux, I.~Tal, A.~Vardy, and W.~J. Gross, ``Hardware architectures for successive cancellation decoding of polar codes,'' in \emph{Proc. 2011 IEEE Int. Conf. Acoust., Speech Signal Process. (ICASSP’11)}, Prague, Czech Republic, May 2011, pp. 1665--1668.

\bibitem{hardware_bp}
Y.~Yan, X.~Zhang, and B.~Wu, ``An implementation of belief propagation decoder with combinational logic reduced for polar codes,'' \emph{IEICE Electron. Express}, vol.~16, no.~15, p. 20190382, Jul. 2019.

\bibitem{hardware_reduced}
------, ``Reduced complexity successive-cancellation decoding of polar codes based on linear approximation,'' \emph{IEICE Trans. Fundam. Electron. Commun. Comput. Sci.}, vol. E103.A, no.~8, pp. 995--999, Aug. 2020.

\bibitem{hardware_scl}
C.~Leroux, A.~J. Raymond, G.~Sarkis, I.~Tal, A.~Vardy, and W.~J. Gross, ``Hardware implementation of successive-cancellation decoders for polar codes,'' \emph{J. Signal Process. Syst.}, vol.~69, no.~3, pp. 305--315, Dec. 2012.

\bibitem{hardware_bch}
P.~Mathew, L.~Augustine, G.~Sabarinath, and T.~Devis, ``Hardware implementation of (63,51) {BCH} encoder and decoder for {WBAN} using {LFSR} and {BMA},'' \emph{arXiv:1408.2908}, Aug. 2014.

\bibitem{pipelined_arch}
S.~Cao, T.~Lin, S.~Zhang, S.~Xu, and C.~Zhang, ``A reconfigurable and pipelined architecture for standard-compatible {LDPC} and polar decoding,'' \emph{IEEE Trans. Veh. Technol.}, vol.~70, no.~6, pp. 5431--5444, Jun. 2021.

\bibitem{unified_fec}
Y.~Yue, T.~Ajayi, X.~Liu, P.~Xing, Z.~Wang, D.~Blaauw, R.~Dreslinski, and H.~S. Kim, ``A unified forward error correction accelerator for multi-mode turbo, {LDPC}, and polar decoding,'' in \emph{Proc. IEEE Int. Symp. Low Power Electron. Des. (ISLPED’22)}, Boston, MA, USA, Aug. 2022, pp. 1--6.

\bibitem{osd_decoder0}
M.~P.~C. Fossorier and S.~Lin, ``Soft-decision decoding of linear block codes based on ordered statistics,'' \emph{IEEE Trans. Inf. Theory}, vol.~41, no.~5, pp. 1379--1396, Sep. 1995.

\bibitem{osd_decoder1}
M.~P.~C. Fossorier, M.~Mihaljevic, and H.~Imai, ``Reduced complexity iterative decoding of low-density parity check codes based on belief propagation,'' \emph{IEEE Trans. Commun.}, vol.~47, no.~5, pp. 673--680, May 1999.

\bibitem{osd_decoder2}
M.~P.~C. Fossorier, ``Iterative reliability-based decoding of low-density parity check codes,'' \emph{IEEE J. Sel. Areas Commun.}, vol.~19, no.~5, pp. 908--917, May 2001.

\bibitem{attn_2017}
A.~Vaswani, N.~Shazeer, N.~Parmar, J.~Uszkoreit, L.~Jones, A.~N. Gomez, and I.~Polosukhin, ``Attention is all you need,'' in \emph{Proc. Adv. Neural Inf. Process. Syst. (NeurIPS’17)}, vol.~30, Long Beach, CA, USA, Dec. 2017.

\bibitem{ecct_2022}
Y.~Choukroun and L.~Wolf, ``Error correction code transformer,'' in \emph{Proc. Adv. Neural Inf. Process. Syst. (NeurIPS’22)}, vol.~35, New Orleans, LA, USA, Nov. 2022, pp. 38\,695--38\,705.

\bibitem{fecct_2024}
------, ``A foundation model for error correction codes,'' in \emph{Proc. 12th Int. Conf. Learn. Represent. (ICLR’24)}, Vienna, Austria, Apr. 2024.

\bibitem{cvpr_2016}
K.~He, X.~Zhang, S.~Ren, and J.~Sun, ``Deep residual learning for image recognition,'' in \emph{Proc. 2016 IEEE Conf. Comput. Vis. Pattern Recognit. (CVPR’16)}, Las Vegas, NV, USA, Jun. 2016.

\bibitem{layer_norm_2016}
J.~L. Ba, J.~R. Kiros, and G.~E. Hinton, ``Layer normalization,'' \emph{arXiv:1607.06450}, Jul. 2016.

\bibitem{dl_decoding_syndrome}
A.~Bennatan, Y.~Choukroun, and P.~Kisilev, ``Deep learning for decoding of linear codes - a syndrome-based approach,'' in \emph{Proc. 2018 IEEE Int. Symp. Inf. Theory (ISIT’18)}, Vail, CO, USA, Jun. 2018, pp. 1595--1599.

\bibitem{capacity_ldpc_mp}
T.~J. Richardson and R.~L. Urbanke, ``The capacity of low-density parity-check codes under message-passing decoding,'' \emph{IEEE Trans. Inf. Theory}, vol.~47, no.~2, pp. 599--618, Feb. 2001.

\bibitem{transformer_long_seq}
H.~Zhou, S.~Zhang, J.~Peng, S.~Zhang, J.~Li, H.~Xiong, and W.~Zhang, ``Informer: Beyond efficient transformer for long sequence time-series forecasting,'' \emph{Proc. AAAI Conf. Artif. Intell.}, vol.~35, no.~12, pp. 11\,106--11\,115, May 2021.

\bibitem{Jensen_Shannon_divergence}
M.~L. Menéndez, J.~A. Pardo, L.~Pardo, and M.~d.~C. Pardo, ``The {Jensen}-{Shannon} divergence,'' \emph{J. Franklin Inst.}, vol. 334, no.~2, pp. 307--318, Mar. 1997.

\bibitem{Kullback_Leibler_divergence}
S.~Kullback and R.~A. Leibler, ``On information and sufficiency,'' \emph{Ann. Math. Stat.}, vol.~22, no.~1, pp. 79--86, Mar. 1951.

\bibitem{hardware_acc}
Y.~Liao, J.~Meng, and J.-s. Seo, ``A 28nm scalable and flexible accelerator for sparse transformer models,'' in \emph{Proc. 29th ACM/IEEE Int. Symp. Low Power Electron. Design (ISLPED '24)}, New York, NY, USA, Jul. 2024, pp. 1--6.

\bibitem{ml_research}
R.~Xiong, Y.~Yang, D.~He, K.~Zheng, S.~Zheng, C.~Xing, H.~Zhang, Y.~Lan, L.~Wang, and T.~Liu, ``On layer normalization in the transformer architecture,'' in \emph{Proc. 37th Int. Conf. Mach. Learn. (ICML’20)}, vol. 119, Virtual Event, Jul. 2020, pp. 10\,524--10\,533.

\bibitem{stochastic_opt}
D.~P. Kingma and J.~Ba, ``Adam: A method for stochastic optimization,'' \emph{arXiv:1412.6980}, Jan. 2017.

\bibitem{database_channel_codes}
\BIBentryALTinterwordspacing
M.~Helmling, S.~Scholl, F.~Gensheimer, T.~Dietz, K.~Kraft, S.~Ruzika, and N.~Wehn, ``Database of channel codes and {ML} simulation results,'' 2019. [Online]. Available: \url{https://www.uni-kl.de/channel-codes}
\BIBentrySTDinterwordspacing

\bibitem{cmp_bp}
H.~E. Kyburg~Jr., ``Probabilistic reasoning in intelligent systems: Networks of plausible inference,'' JSTOR, 1991.

\bibitem{cmp_hyper_bp}
E.~Nachmani and L.~Wolf, ``Hyper-graph-network decoders for block codes,'' in \emph{Proc. Adv. Neural Inf. Process. Syst. (NeurIPS’19)}, vol.~32, Vancouver, BC, Canada, Dec. 2019.

\bibitem{cmp_arbp}
------, ``Autoregressive belief propagation for decoding block codes,'' \emph{arXiv:2103.11780}, Mar. 2021.

\bibitem{ccsds}
CCSDS, ``Recommendation for space data system standards,'' CCSDS, Technical Report, 2003, cCSDS 131.0-B-1, Blue Book.

\bibitem{mackay}
D.~J.~C. MacKay, ``Good error-correcting codes based on very sparse matrices,'' \emph{IEEE Trans. Inf. Theory}, vol.~45, no.~2, pp. 399--431, Mar. 1999.

\bibitem{ldpc_capacity}
T.~J. Richardson and R.~L. Urbanke, ``The capacity of low-density parity-check codes under message-passing decoding,'' \emph{IEEE Trans. Inf. Theory}, vol.~47, no.~2, pp. 599--618, Feb. 2001.

\bibitem{shannon_limit}
C.~E. Shannon, ``A mathematical theory of communication,'' \emph{Bell Syst. Tech. J.}, vol.~27, no.~3, pp. 379--423, Jul. 1948.

\bibitem{ldpc_nms}
J.~Chen and M.~P.~C. Fossorier, ``Decoding low-density parity check codes with normalized {APP}-based algorithm,'' in \emph{Proc. IEEE Global Telecommun. Conf. (GLOBECOM’01)}, vol.~2, San Antonio, TX, USA, Nov. 2001, pp. 1026--1030.

\bibitem{flops_thop}
\BIBentryALTinterwordspacing
L.~Zhu, ``{THOP}: A tool for measuring the {FLOPs} of neural networks,'' 2020. [Online]. Available: \url{https://github.com/Lyken17/pytorch-OpCounter}
\BIBentrySTDinterwordspacing

\bibitem{transformer_papa_share}
S.~Takase and S.~Kiyono, ``Lessons on parameter sharing across layers in transformers,'' \emph{arXiv:2104.06022}, Jun. 2023.

\bibitem{low_rank_attn}
B.~Chen, T.~Dao, E.~Winsor, Z.~Song, A.~Rudra, and C.~Ré, ``Scatterbrain: Unifying sparse and low-rank attention,'' in \emph{Proc. Adv. Neural Inf. Process. Syst. (NeurIPS’21)}, vol.~34, Virtual Event, Dec. 2021, pp. 17\,413--17\,426.

\bibitem{transformer_quantization}
Z.~Liu, Y.~Wang, K.~Han, W.~Zhang, S.~Ma, and W.~Gao, ``Post-training quantization for vision transformer,'' \emph{Adv. Neural Inf. Process. Syst.}, vol.~34, pp. 28\,092--28\,103, 2021.

\end{thebibliography}

\end{document}